\documentclass[aps,twocolumn,groupedaddress,floatfix,superscriptaddress]{revtex4}
\usepackage{amsmath, amssymb, fourier, xcolor}
\usepackage{amsthm}
\usepackage{graphicx}
\usepackage{subfigure}
\usepackage{natbib}
\usepackage[hyperindex]{hyperref}

\newtheorem*{theorem}{Theorem}

\begin{document}
\raggedbottom
\bibliographystyle{apsrev}
\title{A Unified Framework for Cluster Methods with Tensor Networks}

\author{Erdong Guo}
\email{eguo1@ucsc.edu}
\affiliation{University of California, Santa Cruz}
\affiliation{University College London}

\author{David Draper}
\email{draper@ucsc.edu}
\affiliation{University of California, Santa Cruz}

\begin{abstract}
Markov Chain Monte Carlo (MCMC), and Tensor Networks (TN) are two powerful frameworks for numerically investigating many-body systems, each offering distinct advantages.
MCMC, with its flexibility and theoretical consistency, is well-suited for simulating arbitrary systems by sampling.
TN, on the other hand, provides a powerful tensor-based language for capturing the entanglement properties intrinsic to many-body systems, offering a universal representation of these systems.
In this work, we leverage the computational strengths of TN to design a versatile cluster MCMC sampler.
Specifically, we propose a general framework for constructing tensor-based cluster MCMC methods,
enabling arbitrary cluster updates by utilizing TNs to compute the distributions required in the MCMC sampler.
Our framework unifies several existing cluster algorithms as special cases and allows for natural extensions. 
We demonstrate our method by applying it to the simulation of the two-dimensional Edwards-Anderson Model and the three-dimensional Ising Model.
This work is dedicated to the memory of Prof. David
Draper \cite{in_memoriam_david},
\end{abstract}

\maketitle

\textsl{Introduction}:
The Markov Chain Monte Carlo (MCMC) method is a powerful computational tool widely used in the study of Bayesian Statistics 
and many-body mechanical systems \cite{draper1995assessment, davids_notes}.
MCMC offers several advantages, such as flexibility in handling arbitrary systems, consistency, and ease of implementation compared to other numerical methods.
However, MCMC suffers from critical slowing down, which occurs due to the strong autocorrelation of samples near the critical points of the simulated systems.
To address this issue, cluster update methods were developed.
These methods involve updating multiple spin sites within a cluster that percolates through the entire system at the critical point of the system,
significantly alleviating the problem of critical slowing down \citep{swendsen1987nonuniversal, fortuin1972random, wolff1989collective, niedermayer1988general, chayes1997graphical, kent2018cluster, kandel1990cluster, prokof2001worm}.
However, cluster update methods are not universally applicable and are typically limited to specific systems, often relying on the symmetry of the system.
For instance, in cases where the $\mathbf{Z}_{2}$ symmetry of the Ising model is broken, the efficiency of standard S-W type algorithms deteriorates significantly \citep{kent2018cluster}.
Tensor Networks (TN) have shown considerate success in simulating many-body mechanical systems \citep{white1992density, verstraete2004renormalization}.
TNs provide an efficient framework for describing both classical and quantum many-body systems,
capturing the entanglement properties intrinsic to these systems \cite{orus2014practical}.
Several approaches have emerged that combine TN formalism with MCMC and Monte Carlo (MC) methods \citep{ferris2015unbiased, elith2006novel, sandvik2007variational, ferris2012perfect, huggins2017monte, rams2021approximate, ueda2007corner, frias2023collective,
schuch2008simulation,
meurice2014sampling,
stoudenmire2010minimally,
berta2018thermal, nishino1996corner}.
One direction involves sampling from approximated target distributions represented by TN contractions to estimate target quantities,
or using TN-approximated distributions as proposal distributions in Metropolis-Hasting type algorithms \citep{frias2023collective}.
Another category integrates MC methods into Tensor Network Renormalization Group (TNRG) schemes, 
leveraging the benefits of distributed high-performance computing.

In this work, we propose a general framework for constructing cluster MC methods using the language of TN.
This framework enables the reformulation of several existing cluster MC methods,
including the Swendsen-Wang, Wolff, ghost field Swendsen-Wang, Niedermayer, and KBD algorithms.
Specifically, we incorporate auxiliary variables that exploit the structure of tensors,
facilitating the use of Gibbs sampler for simulation through exact conditional distributions computed by tensor contractions.
In scenarios where it is impractical to construct suitable auxiliary variables,
we propose an alternative approach: contracting the cluster of tensors exactly—without introducing auxiliary variables—to obtain the conditional distribution $p(C[\sigma]|\partial C[\sigma])$.
Alternatively, tensors can be contracted approximately to derive an approximated conditional distribution $\tilde{p}(C[\sigma]|\partial C[\sigma])$.
To mitigate the bias introduced by this approximation,
a single step of the M-H is employed, ensuring the target distribution remains the true conditional distribution.
A complete Gibbs sampling step is achieved by iterating this process across all clusters.
A notable special case of this formalism is when the entire spin lattice is treated as one cluster,
leading to the approximated joint distribution $\tilde{p}(\underline{\sigma})$ obtained by contracting the entire tensor network.
This case essentially corresponds to the TNMH algorithm \citep{frias2023collective}.
In summary, TN provides a powerful framework for analytically and numerically computing the conditional distributions of arbitrary clusters, given their boundary configurations,
as well as the joint distribution of the entire lattice configuration.
This capability enables the design of a family of TN-augmented MCMC samplers,
with the Tensor Gibbs (TG) and Tensor Gibbs with M-H (TGMH) formalisms serving as particular implementations within this broader class of methods.

\vspace{1px}
\textsl{Gibbs Samplers Represented by Tensor Networks}:
It is known that statistical mechanical models such as classical Ising model can be formulated as the TNs,
where the partition functions are represented as contractions of tensors defined on corresponding lattices \citep{levin2007tensor}.
This formulation enables the development of MCMC samplers that exploit tensor contraction techniques to compute required probabilities efficiently.
In particular, the conditional distribution $p(C[\sigma]|\partial{C}[\sigma])$ of arbitrary cluster of spins or the joint distribution $p(\sigma_{1:n})$ of the entire spin configuration can be obtained through tensor contractions.
Here, $C[\sigma]$ denotes the spin configuration within the cluster $C$, and $\partial{C}[\sigma]$ refers to the spin configuration on its boundary.
Utilizing these conditional distributions computed via TNs, Gibbs sampling can be employed to simulate the system effectively.
Furthermore, the inherent symmetries of the underlying physical models often result in tensors with favorable structural properties, enabling the introduction of auxiliary variables $\mathbf{h}$.
These auxiliary variables facilitate the decomposition of the TNs in a manner that allows for the exact calculation of the conditional distribution $p(C[\sigma]|\partial{C}[\sigma], \mathbf{h})$ with a closed form.
As we will demonstrate in the subsequent sections, specific constructions of the auxiliary variables $\mathbf{h}$ allow us to derive several well-known cluster MCMC methods,
including the S-W, Wolff, Niedermayer, Ghost Spin S-W, and KBD algorithms.
By considering more general forms of the auxiliary variables,
existing cluster algorithms can be extended straightforwardly,
and the Niedermayer algorithm is generalized following 
this approach as an example.
Since the cluster algorithms derived from TG formalism inherently satisfy the detailed balance condition,
the TG formalism offers a more natural and simpler approach to designing the cluster MCMC methods.
To illustrate the TG formalism, we use two-dimensional Ising model on a square lattice as an example,
although the formalism is applicable to arbitrary systems.

Our target distribution is the Boltzmann distribution $p_{\text{Boltzmann}}(\sigma_{1:n}) = \exp{\{K\sum_{<ij>}\sigma_{i}\sigma_{j}\}} / Z$,
which characterizes the equilibrium of the classical system,
where $Z$ is the partition function (normalization constant).
In this setup, we consider the nearest-neighbor interactions on a square lattice.
In the TG formalism, we introduce another auxiliary variables $\mathbf{h}^{(ij)}$ for each bond connecting sites $i$ and $j$,
in addition to the spin configuration of the lattice $\sigma_{1:n}$.
For each bond $(i, j)$ in the lattice, there exists a Boltzmann matrix $B_{ij}$ that encapsulates the interaction between spins $i$ and $j$.
With these components, we formulate the augmented model incorporating the auxiliary variables $\mathbf{h}$,
\begin{align*}
    &(\mathbf{h}^{(1:k)}|\sigma_{1:n}) \sim p(\mathbf{h}^{(1:k)}|\sigma_{1:n}), \\
    &\sigma_{1:n} \sim p_{\text{Boltzmann}}(\sigma_{1:n}),
\end{align*}
where $\mathbf{h}^{(1:k)}$ represents the $k$ auxiliary variables defined for each edge of the lattice.
We can employ a block Gibbs sampler to iteratively sample both $\sigma_{1:n}$ and $\mathbf{h}^{(1:k)}$ from their respective full-conditional distributions, driving the system toward thermal equilibrium.
Within the TG formalism,
the FK representation \cite{fortuin1972random}, which underlies the well-known S-W and Wolff algorithms \cite{swendsen1987nonuniversal, wolff1989collective}, can be viewed as a specific construction of the conditional distribution $p(\mathbf{h}^{(1:k)}|\sigma_{1:n})$.
This construction allows for the analytical derivation of full-conditional distributions for both $\mathbf{h}$ and $\sigma_{1:n}$ enabling efficient sampling.
Compared to the FK representation, the tensor network approach offers a systematic method for computing these full-conditional distributions.
As an example of the TG formalism in action,
we can construct a more general auxiliary variable $\mathbf{h}^{(1:k)}$ by exploiting the structure of the tensors $B_{ij}$,
leading to the analytic derivation of full-conditional distributions.
This resulting method naturally extends Niedermayer's algorithm.

\begin{figure}[ht]
\centering
\subfigure{
    \includegraphics[width=0.48\columnwidth]{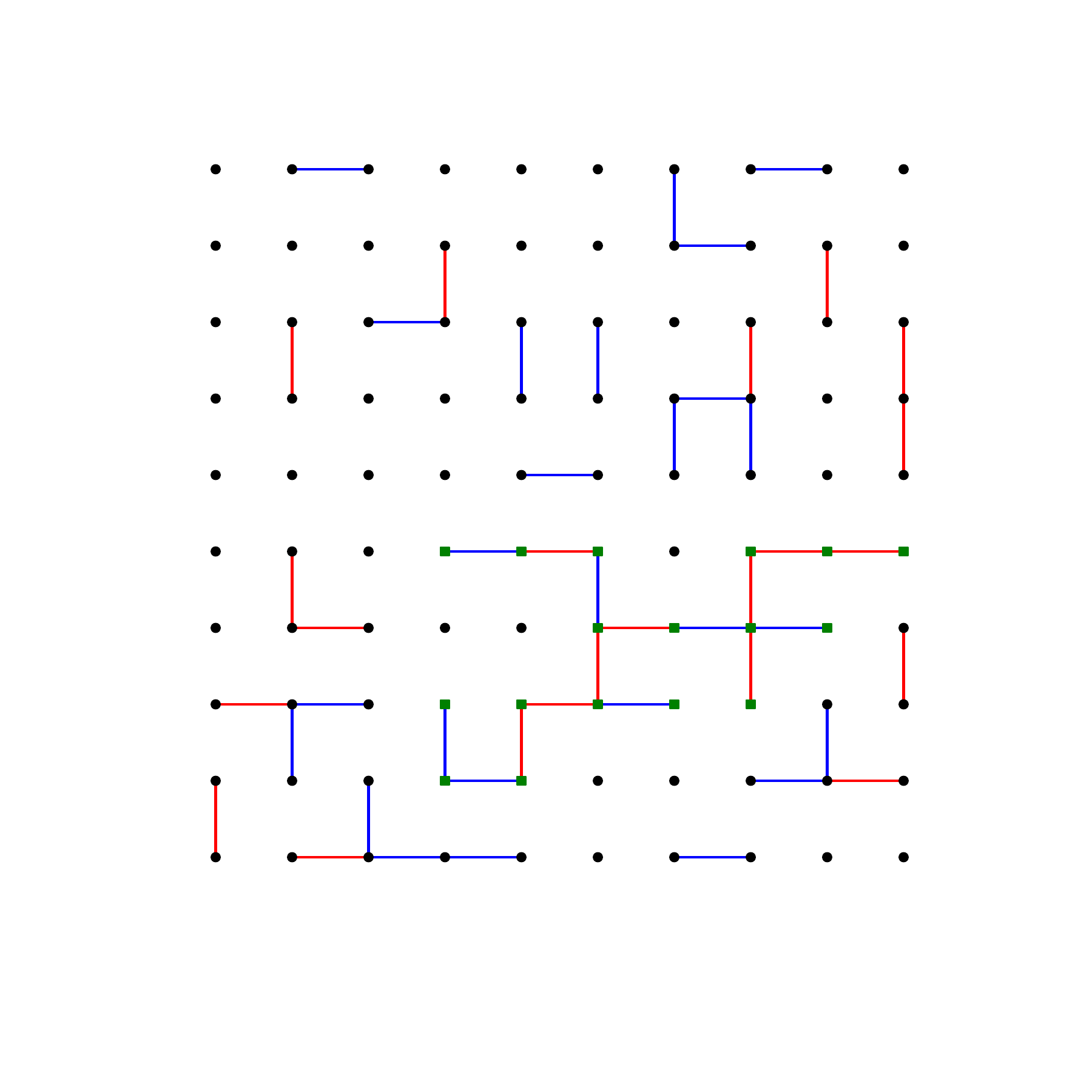}
    \label{fig: tsw_full}
}
\subfigure{
    \includegraphics[width=0.48\columnwidth]{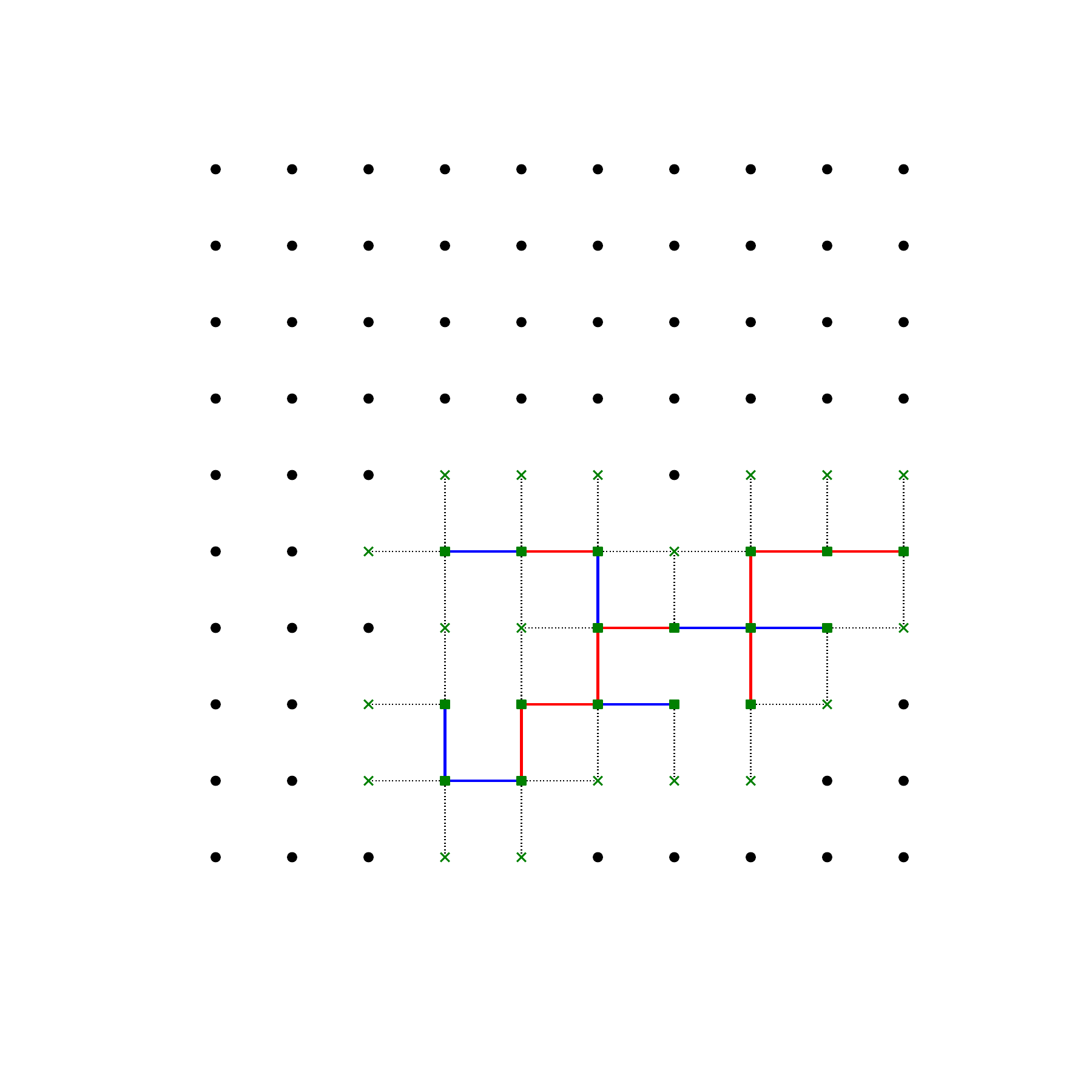}
    \label{fig: tsw_one_component}
}
\caption{(Left) Illustration of the sampling process for the auxiliary variables $\mathbf{h}$ in the TG sampler. Edges are categorized into three types: type 1 (red), type 2 (blue), and type 3 (empty), each representing different bond states.
(Right) Depiction of a specific cluster with boundary spins treated as external fields. Dashed edges represent the external field vectors, which describe interactions between the boundary spins and the spins within the cluster.}
\label{fig: tg_gibbs}
\end{figure}

We introduce the auxiliary variable $h_{l}^{(ij)}$ to establish the following relationship between the Boltzmann matrix and bond state variable:
\begin{equation}
\label{eq: boltzmann}
    B_{ij} = \sum_{l=1}^{3}B^{l}_{ij},
\end{equation}
where
$B_{ij}^{1} = 
\begin{pmatrix}
    e^{K} - X & 0 \\
    0 & e^{K} - X
\end{pmatrix}$,
$B_{ij}^{2} = 
    \begin{pmatrix}
    0 & e^{-K} - Y \\
    e^{-K} - Y & 0 
    \end{pmatrix}$,
and
$B_{ij}^{3} = 
\begin{pmatrix}
   X & Y \\
   Y & X
\end{pmatrix}$,
and $X \in [0, e^{K}]$, and $Y \in [0, e^{-K}]$,
ensuring a valid probability explanation.
Consequently, the distribution of the bond state variable $h_{l}^{(ij)}$ for bond $(i, j)$ can be derived as
\begin{align}
\label{eq: bond_conditional_prob}
    &(h^{(ij)}_{l}|\sigma_{i}, \sigma_{j}) \sim \text{Mult}(h^{(ij)}_{l}|1, \mathbf{p}(\sigma_{i}, \sigma_{j})),
\end{align}
where $\mathbf{p}(\sigma_{i}, \sigma_{j}) = (p_{1}, p_{2}, p_{3})$, $p_{1} = (1 - Xe^{-K})\delta_{\sigma_{i}, \sigma_{j}}$, $p_{2} = (1 - Ye^{K})\delta_{\sigma_{i}, -\sigma_{j}}$ and $p_{3} = Xe^{-K}\delta_{\sigma_{i}, \sigma_{j}} + Ye^{K}\delta_{\sigma_{i}, -\sigma_{j}}$.
$\text{Mult}(\cdot|1, \mathbf{p})$ denotes the multinomial distribution with $1$ trial, three categories, and probability vector $\mathbf{p}(\sigma_{i}, \sigma_{j})$.
This means that the bond state variable $h^{(ij)}_{l}$ for bond $(i, j)$ is a one-hot vector,
taking one of the three possible states with a probability vector $\mathbf{p}$ that depends on the spin configuration at the bond's endpoints $i$ and $j$.
Due to this construction, the variables $\mathbf{h}^{1:k}$ are conditionally independent given the spin configurations at the corresponding bond endpoints.
After the bond states $h^{1:k}_{l}$ are sampled from the conditional distributions given the current spin configurations,
as illustrated in Fig. \ref{fig: tsw_full}, the tensor network representing the current spin lattice simplifies.
It now consists of three types of matrices
$B^{1}_{ij}$, $B^{2}_{ij}$ and $B^{3}_{ij}$
corresponding to the blue, red, and empty edges in the figure.
With the aforementioned simplification,
the full-conditional distribution $p(\sigma_{1:n}|\cdot)$ of the spin configuration can be analytically calculated by contracting the resulting tensors.
As depicted in Fig. \ref{fig: tsw_one_component},
if we focus on the conditional distribution of a cluster of spins connected by Boltzmann matrices of type $1$ and $2$,
we only need to contract the matrices within the cluster while treating the spins at its boundary as external fields.
The resulting conditional probability ratio of the only two non-zero cluster configurations is then expressed as
\begin{align}
    \frac{p(C_{i}[\sigma]|C_{i}[\mathbf{h}], \partial{C_{i}}[\sigma])}{p(C_{i}[\bar{\sigma}]|C_{i}[\mathbf{h}], \partial{C_{i}}[\sigma])} = (\frac{X}{Y})^{n-m},    
\label{eq: master_evv}
\end{align}
where $\bar{\sigma}$ denotes the configuration of the cluster with all spins flipped, and $n$ and $m$ denotes the number of boundary edges connecting the parallel and anti-parallel spin pairs respectively, as shown by dashed edges in Fig. \ref{fig: tsw_one_component}.

The detailed derivation of the above formula by tensor contraction is provided in Appendix \ref{app: contraction_cond}.
However, this result can be understood intuitively: 
because only diagonal and anti-diagonal matrices exist within the cluster, there are only two possible spin configuration, as the spins are "locked" together, preventing individual flip.
We summarize one step of the TG sampler algorithm as follows: 
\begin{itemize}
    \item Sample auxiliary variables:
    For each edge $(i, j)$, sample the auxiliary variable $h^{(ij)}_{l}$ from the corresponding conditional distribution $p(h^{(ij)}_{l}|\underline{\sigma})$;
    \item Update spin configurations: For all clusters, compute the conditional distribution of the spin configuration of the cluster $C_{i}$ by contracting the resulting tensors of the cluster. Update the spin configuration by sampling from this conditional distribution.
\end{itemize}
In addition to direct sampling, the M-H method can be applied to sample from the conditional distributions.
According to Eq. \ref{eq: master_evv}, the conditional probability ratio of the cluster configuration is $1$ when $X = Y = e^{-K}$.
By flipping all spins in the cluster with a probability of $0.5$, we recover S-W algorithm.
However, if we apply the M-H method with a deterministic flip proposal for the entire cluster configuration,
the cluster spins will flip together with probability $1$, yielding the Wolff algorithm.

Not only are the S-W and Wolff algorithms special cases of the TG sampler, 
but it is also straightforward to show that Niedermayer's algorithm \cite{niedermayer1988general} is a specific implementation of the TG formalism.
As presented in Eq. \ref{eq: inclusion_tune}, a tunable parameter $W$ is introduced in the bond inclusion probability of Niedermayer's algorithm,
aiming to control the average cluster size and thus minimize the relaxation time.
\begin{equation}
\label{eq: inclusion_tune}
    P_{\text{add}}(h^{(ij)}|\sigma_{i}, \sigma_{j}) = 
    \begin{cases}
    1 - We^{-K\sigma_{i}\sigma_{j}} & \text{if } W < e^{K\sigma_{i}\sigma_{j}}, \\
    0 & \text{otherwise}.
    \end{cases}
\end{equation}
To recover Niermayer's algorithm from the TG algorithm,
specific relationships betweeen $X$ and $Y$ in Eq. \ref{eq: bond_conditional_prob} can be set, as detailed in Table. \ref{tab: ours_vs_niedermayer}.
Notably, when $W$ lies within the range $[0, e^{-K})$, the type $1$ and type $2$ clusters are combined in Niedermayer's method as derived from the TG formalism.
This indicates that the TG formalism allows for different bond inclusion probabilities for parallel and anti-parallel spin clusters. 
\begin{table}[]
\vspace{2mm}
\centering
\begin{tabular}{lc|l}\hline\hline
\multicolumn{2}{c|}{TG Framework} & Niedermayer's \\ \hline
\multicolumn{1}{c}{$X$} &
\multicolumn{1}{c|}{$Y$} &
\multicolumn{1}{c}{$W$} \\ \hline
$X \in [0, e^{-K})$ & $X$ & $W \in [0, e^{-K})$ \\
$X \in [e^{-K}, e^{K}]$ & $Y = e^{-K}$ & $W \in [e^{-K}, e^{K}]$\\
\hline\hline
\end{tabular}
\caption{The mapping from Niedermayer's method to special case of TG method by constraining parameters $X$ and $Y$ in particular ranges.}
\label{tab: ours_vs_niedermayer}
\end{table}

\begin{figure}[ht]
\centering
\subfigure{
    \includegraphics[width=0.48\columnwidth]{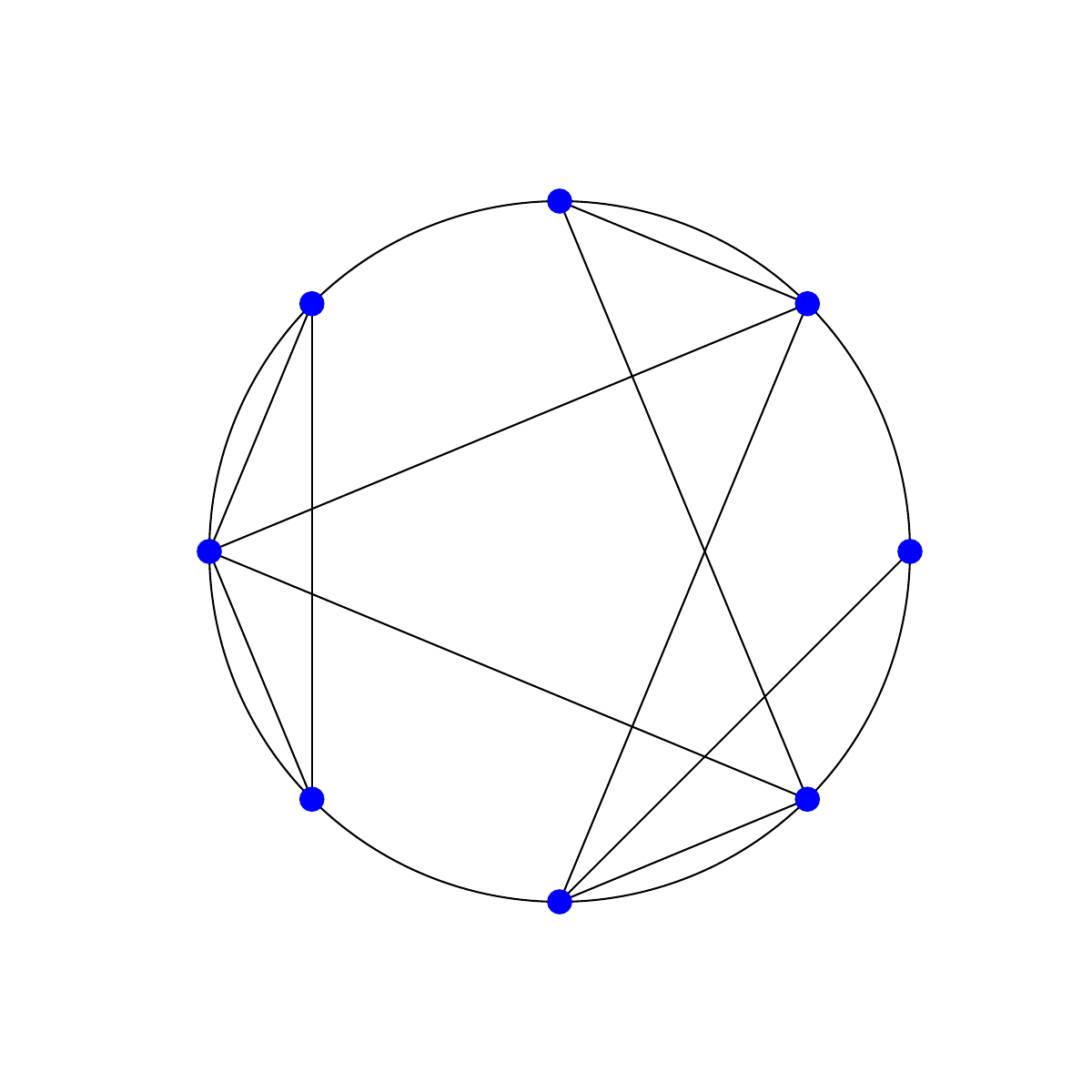}
    \label{fig: sw_ng}
}
\subfigure{
    \includegraphics[width=0.48\columnwidth]{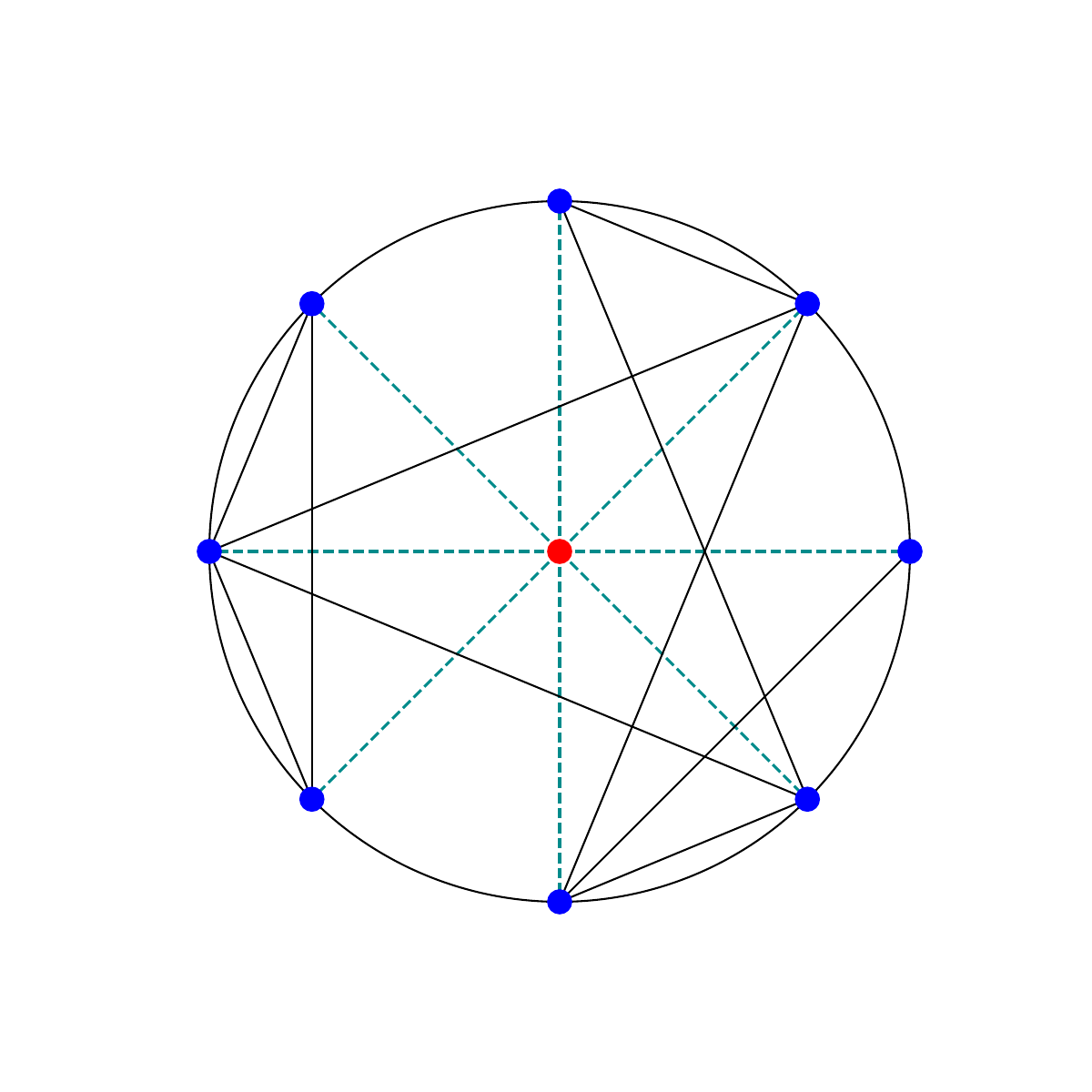}
    \label{fig: sw_g}
}
\caption{Figure (left) shows the tensor networks representation of a spin lattice system with external field.
Figure (right) shows augmented tensor networks with ghost tensor denoted by red point, and the blue point denotes the normal tensor of the normal spin. Dark solid edge represents the Boltzmann matrix $B_{ij}$ and dashed green edge represents the Boltzmann matrix $C_{ij}$.}
\label{fig: swg}
\end{figure}

\vspace{1px}
\textsl{Dealing with External Field:}
As TG Sampler is versatile and can be applied to arbitrary systems,
we briefly outline its application to the two-dimensional square lattice model with an external field. 
For a general square lattice model with external fields, where the energy function is given by
$H = -J\sum_{<ij>}\sigma_{i}\sigma_{j} - B_{i}\sigma_{i}$,
Although the external field interacts with each site, it does not alter the distribution of the auxiliary variable.
However, it does impact the full-conditional distribution of the cluster configuration, as the external field modifies the tensors associated with the cluster.
Similar to the derivation process introduced in Appendix \ref{app: contraction_cond},
the probability ratio is obtained as
\begin{align}
    \label{eq: master_evv_ea}
    \frac{p(C_{i}[\sigma]|C_{i}[\mathbf{h}], \partial{C_{i}}[\sigma])}{p(C_{i}[\bar{\sigma}]|C_{i}[\mathbf{h}], \partial{C_{i}}[\sigma])} = (\frac{X}{Y})^{n-m}e^{2\tilde{B}(p - q)},
\end{align}
where $\tilde{B} = \beta B$, and $p$ and $q$ represent the numbers of spin that are parallel and anti-parallel to $B$ in $C_{i}(\sigma)\cup\partial[\sigma]$, respectively.
As indicated by the above formula, the external field term dominates the probability ratio at low temperatures, causing the cluster configuration to align with the external field.
Theoretically, this phenomenon occurs because the external field breaks the $\mathbf{Z}_{2}$ symmetry of the system, which disrupts the cluster algorithm.
To address this issue, a novel approach involves introducing an additional variable known as the "ghost spin" to restore the system's symmetry without altering the expected measurements of the original system \cite{kent2018cluster}.
In the framework of TNs, the "ghost spin" is represented as an auxiliary tensor, which is a function of the symmetry group elements of the simulated system.
This tensor is integrated into the original lattice and connected with all other tensors located at each site of the spin lattice, as illustrated in Fig. \ref{fig: swg}.

For the augmented TN with "ghost" tensor describing the two-dimensional square lattice with an external field,
the edges connecting the normal spins are represented by the Boltzmann matrix $B_{ij}$ as Eq. \ref{eq: boltzmann}.
The Boltzmann matrix $C_{ij}$ for the edge connecting a normal spin $i$ and a "ghost" spin $j$ is as
\begin{align}
    C_{ij} =
    \begin{pmatrix}
    e^{\tilde{B}} & e^{-\tilde{B}}\\
    e^{-\tilde{B}} & e^{\tilde{B}}
    \end{pmatrix},
\end{align}
where $\tilde{B} = \beta B$. Here, $\sigma_{\text{ghost}} \in \{+1, -1\}$ is evaluated at the elements of $\mathbf{Z}_{2}$ symmetry group.  
Given this tensorized representation of the spin lattice system with an external field, we can apply the TG formalism for simulation.
Specifically, similar to the auxiliary variable introduction strategy as described in Eq.\ref{eq: boltzmann},
we can decompose $C_{ij} = \sum_{l}C^{l}_{ij}$,
where $h$ denotes the "cluster" state to which the edge $(i, j)$ belongs.
This construction effectively removes the external field from the conditional probability ratio of the cluster configuration,
allowing clusters to be flipped efficiently, analogous to the S-W type method.
Correspondingly, the measurement $\mathbb{E}[M]$ of quantity $M$ is defined as $\mathbb{E}[M(\sigma_{\text{ghost}}*\sigma_{1:n})]$.
More detailed illustrations of this construction can be found in Appendix \ref{app: ghost_tensor}.

\vspace{1px}
\textsl{Tensorized KBD algorithm:}
The KBD algorithm was proposed to effectively simulate fully frustrated Ising (FFI) models on a square lattice with nearest-neighbor interactions,
as traditional cluster algorithms, such as S-W, Wolff and their variants become less effective when dealing with models characterized by competing interactions and frustration \cite{elith2006novel, kandel1992cluster}.
Numerical simulations in \cite{elith2006novel, kandel1992cluster} have demonstrated that while the single spin flip M-H algorithm suffers from severe critical slowing down, 
the KBD algorithm remains efficient even at zero temperature.

Although the construction of bond and cluster-flipping distributions in KBD satisfies detailed balance—thereby ensuring the algorithm evolves the system toward equilibrium—the underlying rationale for this construction is not immediately apparent, and it remains unclear whether a fundamental principle guides these designs.

From the perspective of TN,
the KBD algorithm operates by contracting the four Boltzmann matrices, $B_{ij}$, at each plaquette to create a "coarse grained" TN, where the basic unit tensors are four-way tensors $T_{mnop}$.
When applying the TG formalism to these "coarse grained" TNs,
auxiliary variables $\mathbf{h}^{(i)}$ are introduced by exploiting the symmetry inherent in the resulting four-way tensors,
making the contractions of the resulting TNs analytically tractable.
Compared to S-W type cluster methods,
which use the Boltzmann matrix $B_{ij}$ as the basic tensor unit,
the contraction of tensors within each plaquette in the KBD algorithm results in
the homogeneous coarse-grained TNs.
These capture correlations between bonds through the four-way tensors $T_{ijkl}$.

\begin{figure}[ht]
\centering
\subfigure{
    \includegraphics[width=0.48\columnwidth]{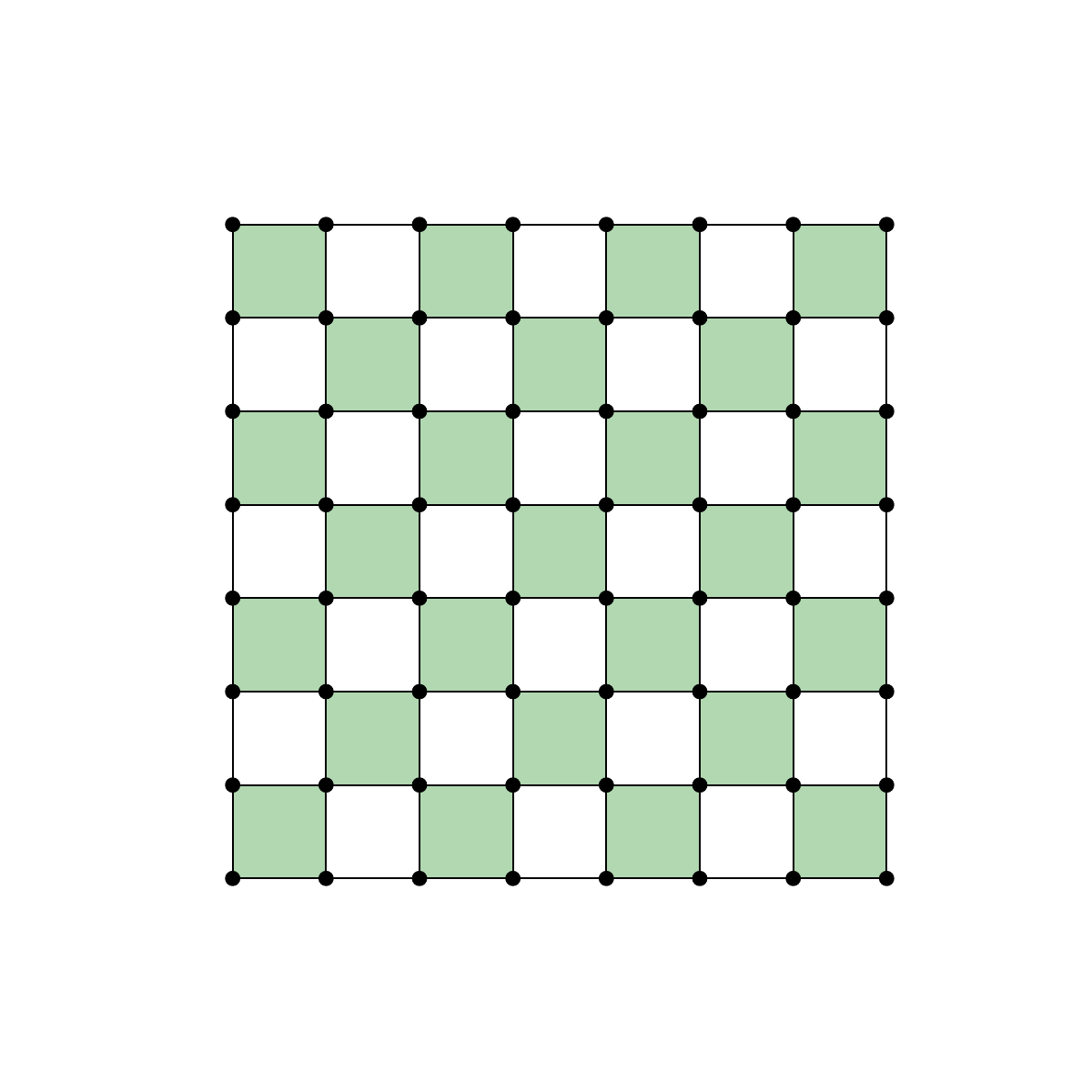}
    \label{fig: kbd_plaquette}
}
\subfigure{
    \includegraphics[width=0.48\columnwidth]{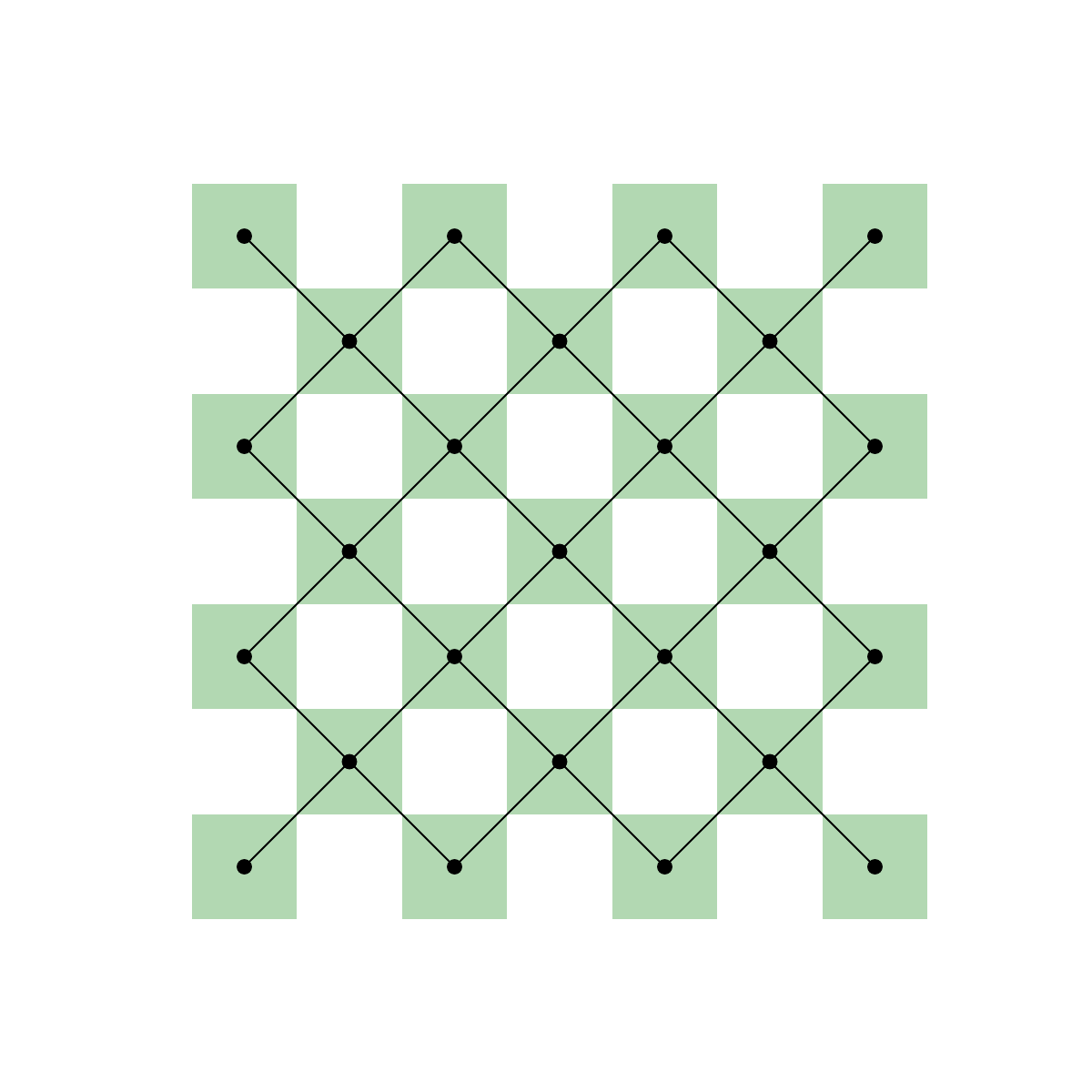}
    \label{fig: kbd_tensor}
}
\caption{(Left) illustrates the 2D FFI model on a square lattice, represented by tensor networks with Boltzmann matrices as block tensors. Each green box highlights a region with frustration.
(Right) depicts the augmented tensor networks, with block tensors $T_{mnop}$ located at the center of each green box.}
\label{fig: kbd}
\end{figure}

In the case of FFI models, each plaquette is frustrated, containing three ferromagnetic bonds and one antiferromagnetic bond, as illustrated in Fig. \ref{fig: kbd_plaquette}.

Without loss of generality, we assume a specific configuration of couplings within a plaquette, 
which leads to the four-rank tensor $T^{(i)}_{mnop}$ corresponding to the contraction of the four Boltzmann matrices located at each plaquette. 
\begin{align}
    T^{(i)}_{mnop} = 
    \begin{pmatrix}
    \begin{pmatrix}
    e^{2K} & e^{-2K} \\
    e^{2K} & e^{2K} 
    \end{pmatrix}
    & \begin{pmatrix}
    e^{2K} & e^{-2K} \\
    e^{-2K} & e^{-2K} 
    \end{pmatrix} \\
    \begin{pmatrix}
    e^{-2K} & e^{-2K} \\
    e^{-2K} & e^{2K} 
    \end{pmatrix} 
    &\begin{pmatrix}
    e^{2K} & e^{2K}\\
    e^{-2K} & e^{2K}
    \end{pmatrix}
    \end{pmatrix},
\end{align}
The coarse-grained TN, shown in Fig. \ref{fig: kbd_tensor}, is composed of these tensors.
The auxiliary variables $h_{l}$ and tensor $T_{mnop}$ are related by the following expression:
\begin{align}
    T_{mnop} = \sum_{l}T^{l}_{mnop},
\end{align}
where $T^{1}_{mnop} = e^{2K}\delta{(\sigma_{m}, \sigma_{n})}\delta{(\sigma_{o}, \sigma_{p})}$,
$T^{2}_{mnop} = e^{2K}\delta{(\sigma_{n}, \sigma_{o})}\delta{(\sigma_{m}, -\sigma_{p})}$,
$T^{3}_{mnop} = e^{-2K}v_{n}\otimes v_{o}\otimes v_{m}\otimes v_{p}$,
and $v=(1, 1)$.
Through this construction, the processed TNs,
obtained by contracting the TNs with the observed auxiliary variables $h_{l}$,
can be analytically contracted to provide a closed-form expression for the conditional distribution $p(C[\sigma]|\partial C[\sigma], \mathbf{h}^{(1:k)})$, 
enabling the application of Gibbs sampler to evolve the system.
Moreover, with the same reasoning for TG formalism of S-W type methods,
sampling from the computed conditional distribution $p(C[\sigma|\partial C[\sigma], \mathbf{h}^{(1:k)})$ involves flipping the spins of the entire cluster with a probability $0.5$ or $1.0$, depending on whether the S-W type or Wolff type is used.
We note the three tensors: $T^{1}_{mnop}$, $T^{2}_{mnop}$, and $T^{3}_{mnop}$ encode the bond configuration information as proposed in the KBD algorithm.
Specifically, for the auxiliary variable $h_{l} = (1, 0, 0)$ leading to $T^{1}_{mnop}$, 
the bond between spin pairs $(m, n)$ and $(o, p)$ is frozen, while the bond between spin pairs $(m, p)$ and $(n, o)$ is removed.
For the auxiliary variable $h_{l} = (0, 1, 0)$, leading to $T^{2}_{mnop}$,
the bond between spin pair $(n, o)$ and $(m, p)$ is frozen,
with the bond between spin pair $(m, n)$ and $(o, p)$ removed.
The last state, $h_{l}=(0, 0, 1)$ corresponds to tensor $T^{3}_{mnop}$, indicating that the probabilities for all possible spin configurations of the plaquette are equivalent. 

In summary, the design of auxiliary variables in the KBD algorithm follows a similar rationale to the S-W type algorithm from the TG perspective, by decomposing the original tensor $T_{mnop}$ into "diagonal", "anti-diagonal", and "uniform" tensors, where all components are identical. 
While the TG formalism can naturally extend the KBD algorithm by introducing parameters into the three coarse-grained tensors $T^{1}$, $T^{2}$ and $T^{3}$,
exploring these extensions lies beyond the scope of this work.
However, the trick of simplifying the TNs by auxiliary variables may not be applicable to general models that lack inherent symmetries,
such as the well-known Edwards-Anderson (E-A) models.
Specifically, for systems without clear symmetries that can be leveraged to decompose the TNs in a manner allowing for analytical contraction,
we incorporate the M-H into the TG formalism.
This integration enables unbiased sampling utilizing the approximated conditional distribution $\tilde{p}(C[\sigma]|\partial C[\sigma])$ computed through numerical tensor contractions \citep{frias2023collective}.

\begin{figure}[ht]
\centering
\subfigure[]{
    \includegraphics[width=0.45\columnwidth]{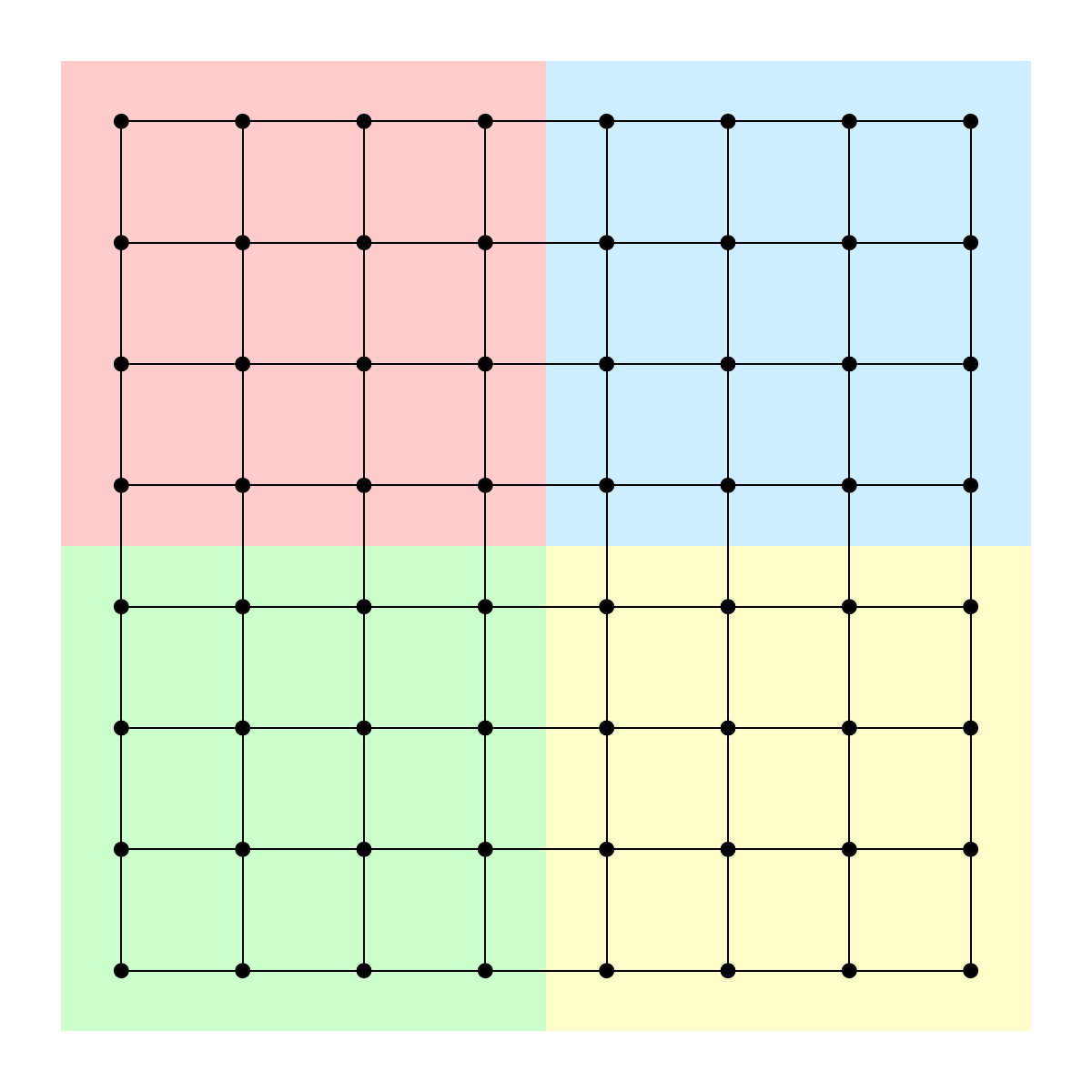}
    \label{fig: tgmh_clusters}
}
\subfigure[]{
    \includegraphics[width=0.49\columnwidth]{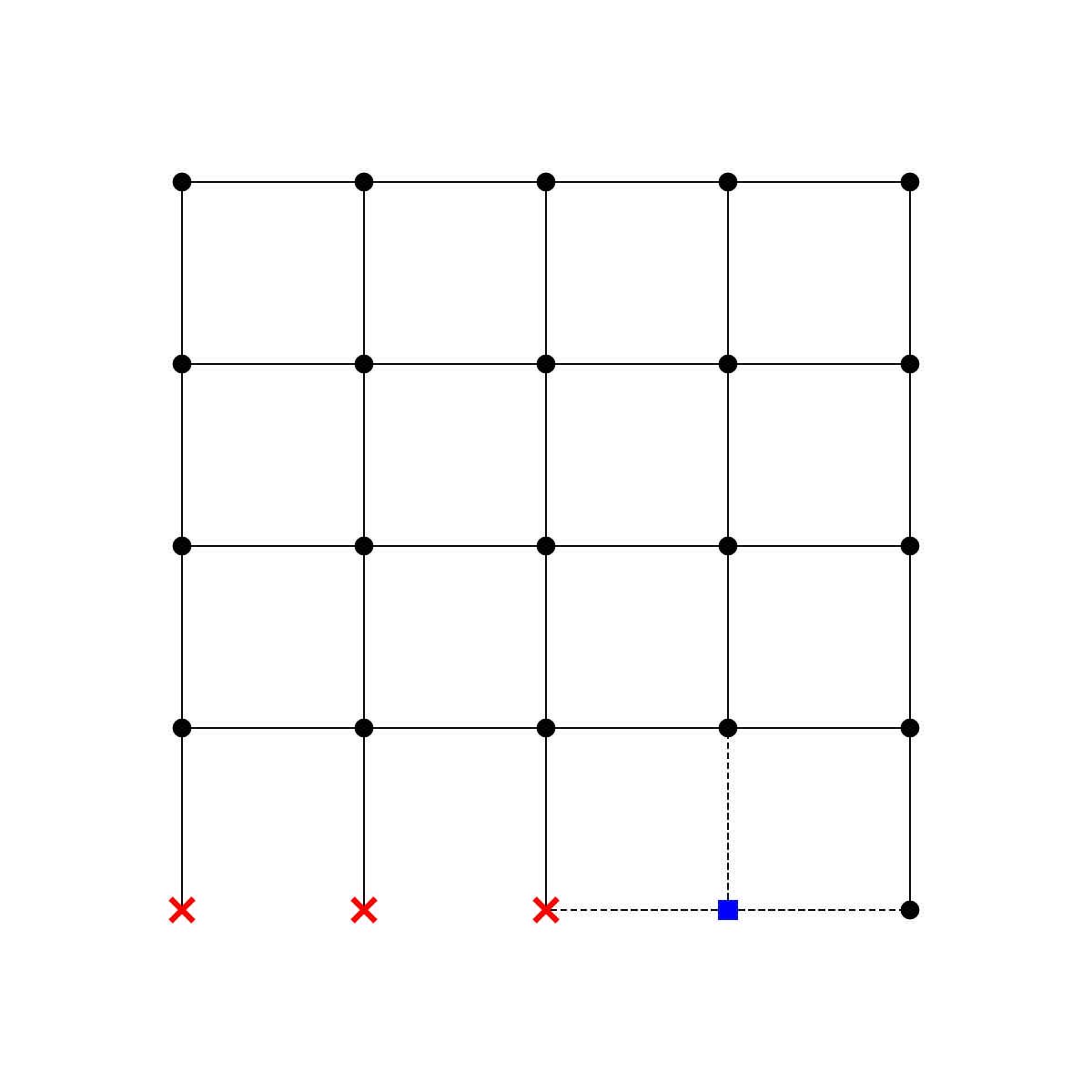}
    \label{fig: tgmh_sampling}
}
\caption{
(a) depicts an example of a spin lattice decomposition into clusters, which are distinguished by different colors.
The TGMH framework supports arbitrary cluster decompositions.
(b) shows the computation of the proposal distribution $\tilde{p}(\sigma_{4}|\sigma_{1:3}, \partial{C}[\sigma])$, where the previously sampled $\sigma_{1:3}$ as well boundary spins $\partial{C}[\sigma]$ are treated as external fields.}
\label{fig: tgmh}
\end{figure}

\vspace{1px}
\textsl{Tensor Gibbs with Metropolis-Hasting (TGMH): }
For the general models lacking symmetries suitable for constructing auxiliary variables,
we contract the tensor networks representing a cluster of the lattice numerically without introducing auxiliary variables,
while appropriately handling the boundary tensors,
to obtain an approximated conditional distribution $\tilde{p}(\text{C}(\sigma)|\partial{\text{C}}[\sigma])$.  
To mitigate the bias introduced by bond truncation in the tensor contraction,
a M-H type acceptance-rejection step is incorporated to target the exact conditional distribution $p(C[\sigma]|\partial C[\sigma])$ of the cluster's spin configuration given the boundary spins. 
Due to the flexibility in computing the approximated conditional distribution of cluster spin configuration,
the TGMH sampler supports updates of arbitrary clusters as shown in Fig. \ref{fig: tgmh},
unlike the TG sampler,
where cluster formation is dictated by auxiliary variables.
As is known, the correlation length in the statistical mechanical systems typically diverges around their critical points,
making larger cluster updates beneficial for reducing the autocorrelation of samples.
Ideally, when the entire lattice is chosen as a single cluster,
the TGMH formalism recovers to TNMH sampler \citep{frias2023collective},
where the approximated join distribution $\tilde{p}(\sigma_{1:n})$ is obtained by contracting all tensors for the entire lattice,
rendering the Gibbs step unnecessary.
Another noteworthy special case occurs with small lattice,
where the joint distribution of the entire lattice can be computed by exactly contracting the tensor networks.
This leads to an M-H acceptation probability of $1$, 
resulting in a perfect sampler where all samples precisely follow the target distribution and are completely independent with each other.

To elucidate the TGMH scheme in detail,
we consider the two-dimensional square lattice as an example,
although this scheme is general and applicable to arbitrary systems.
Let the spins in a cluster $C[\sigma]$ be denoted as $\sigma_{1:k}$ in the following context.
The goal is to generate samples from the proposal distribution $\tilde{p}(\sigma_{1:k}|\partial C[\sigma])$ defined by the tensor networks on the square lattice.
The technique is to sequentially sample the spins $\sigma_{i}$ one at a time from the approximated conditional distribution $\tilde{p}(\sigma_{i}|\sigma_{1:i-1}, \partial{C}[\sigma])$.
Once all spins $\sigma_{1:k}$ in the cluster have been iterated through,
a complete sample $\sigma_{1:k}$ is obtained along with its corresponding probability under the proposal distribution,
which can be computed using the chain rule: $\tilde{p}(\sigma_{1:k}|\partial C[\sigma]) = \tilde{p}(\sigma_{1}|\partial C[\sigma])\prod_{i=2}^{k}\tilde{p}(\sigma_{i}|\sigma_{1:i-1}, \partial C[\sigma])$).
With above scheme to compute the proposal distribution $\tilde{p}(\cdot)$ for an arbitrary spin configuration,
the M-H acceptance probability for the proposed sample can be simply expressed as
\begin{align*}
    1 \wedge \left\{e^{\beta(H[\sigma_{1:k}, \partial C[\sigma])] - H[\sigma^{\prime}_{1:k}, \partial C[\sigma])])}\frac{\tilde{p}(\sigma_{1:k}|\partial C[\sigma]))}{\tilde{p}(\sigma^{\prime}_{1:k}|\partial C[\sigma]))}\right\},
\end{align*}
where $\sigma_{1:k}$ and $\sigma^{\prime}_{1:k}$ denote the current and proposed states for the cluster $C$, individually, and $H(\sigma_{1:k}, \partial C[\sigma]))$ represents the energy of the subsystem consisting of the cluster spins and the boundary spins under simulation. 

As shown in Fig. \ref{fig: tnmc}, we compute the conditional distribution $\tilde{p}(\sigma_{i}|\sigma_{1:i-1}, \partial C[\sigma]))$ row by row,
leveraging the efficient contraction method of Matrix Product States (MPS) for the square lattice, such as the boundary MPS methods. 
To be more specific, from Fig. \ref{fig: tnmc_1} to Fig. \ref{fig: tnmc_4},
the sampling proceeds row by row,
moving from the bottom to the top of the lattice.
The blue point indicates the spin currently being sampled,
and the tensor with dashed lines located at the blue square represents the sampling tensor with one additional index,
which preserves the probability weights of both the up and down states of the sampling spin,
rather than summing them into a scalar.
The red crosses represent external field tensors (vectors) contributed by the observed spins, including the sampled spins $\sigma_{1:i-1}$ as well as the boundary spins $\partial{C}[\sigma]$ of the cluster,
contribute only a specific component (either up or down) to the partition function,
effectively acting as external fields.
A more detailed illustration of this construction is provided in the Appendix \ref{app: tncm_tensors}.
With the above updates to the tensor networks, reflecting the current lattice state,
we can contract all tensors, with one index $l$ remaining introduced by sampling tensor, to obtain the (potentially unnormalized) probability vector
$\tilde{p}^{l} = [\tilde{p}(\sigma_{i}=1|\sigma_{1:i-1}, \partial C[\sigma]), \,\tilde{p}(\sigma_{i}=-1|\sigma_{1:i-1}, \partial C[\sigma])]$.

\vspace{1px}
\textsl{Experiments: $2D$ Edwards-Anderson Model.}
In this section, we investigate the two-dimensional ($\pm{J}$) Edwards-Anderson model with open boundary condition by the TNPC method.
The primary observables analyzed are the energy $H$ and the spin overlap $q$ defined as 
$q=|\frac{1}{n}\sum_{i}\sigma^{(1)}_{i}\sigma^{(2)}_{i}|$, 
where $\sigma^{(1)}_{i}$ and $\sigma^{(2)}_{i}$ denote the $i$-th spin from two independent realizations of the system with the same coupling configuration \cite{young1983statics, bhatt1988numerical, wang1988low}.

\begin{figure}[ht]
\centering
\subfigure[]{
    \includegraphics[width=0.49\columnwidth]{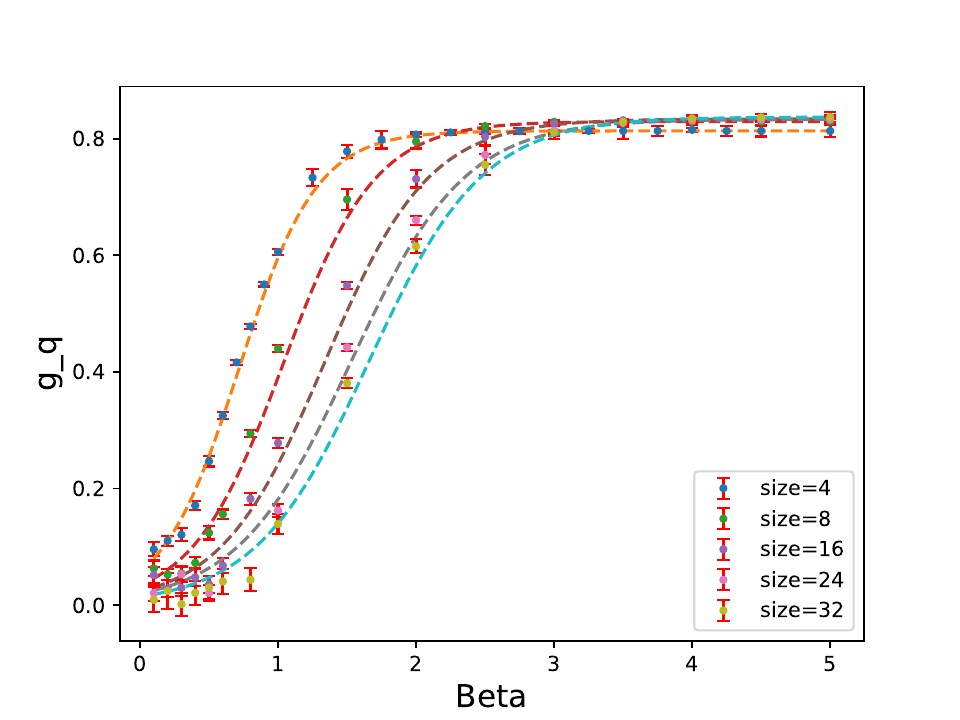}
    \label{fig: g_q_plot}
}
\subfigure[]{
    \includegraphics[width=0.44\columnwidth]{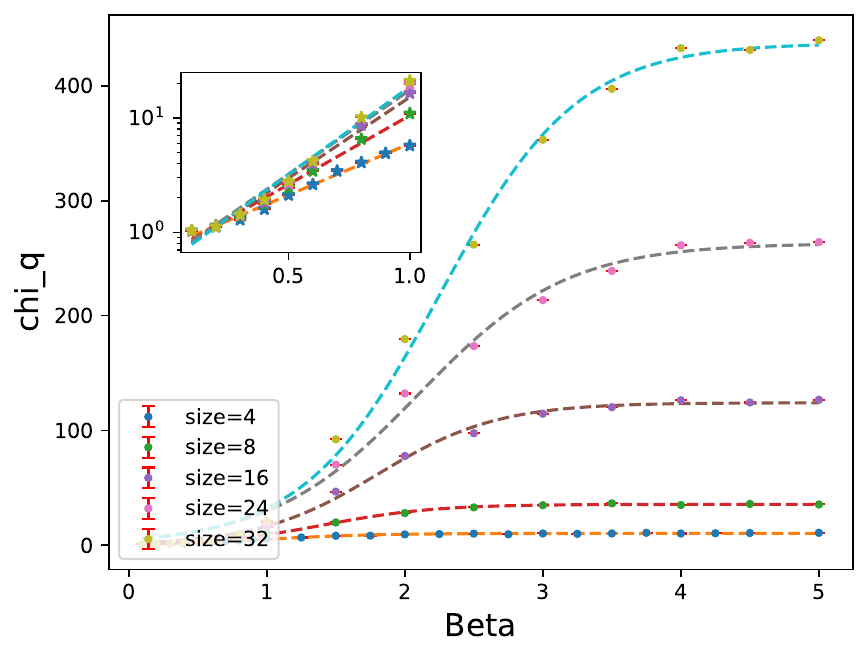}
    \label{fig: chi_q_plot}
}
\caption{(Left) $g(q)$ of the 2D E-A model as a function of $\beta$ for various lattice sizes $L=4, 8, 16, 24, 32$. The curves highlight the trend of crossing shifting with increasing lattice size.
(Right) $\chi(q)$ of the 2D E-A model as a function of $\beta$ for different lattice sizes. The inset provides an enlarged view of the high-temperature region.}
\label{fig: stats_q_plot}
\end{figure}

In the experiments, we vary the temperature over a range corresponding to inverse temperatures from $\beta=0.1$ to $\beta=5.0$, with lattice sizes $L=4, 8, 12, 16, 32$.
We perform disorder averaging over $200$ realizations for all lattice sizes except for the linear size of $32$, where $100$ disorder realizations are used. 
Each simulation involves $10$ burn-in steps followed by $100$ sampling steps.
We analyze the Binder ratio $g(q)$ and the susceptibility $\chi(q)$ of the spin overlap $q$, defined as 
\begin{align}
&g(q) = \frac{1}{2}\langle 3 - \frac{\langle q^{4} \rangle_{\text{T}}}{\langle q^{2} \rangle^{2}_{\text{T}}}\rangle_{\text{J}}, \\
&\chi(q) = N\langle\langle q^{2} \rangle_{\text{T}}\rangle_{\text{J}}
\end{align}
where $\langle\cdot\rangle_{\text{T}}$ and $\langle\cdot\rangle_{\text{J}}$ represent thermal and disorder averages, respectively.
Fig. \ref{fig: stats_q_plot} illustrate how both the Binder ratio and the susceptibility of the spin overlap vary with respect to $\beta$ at different sizes.

As shown in Fig. \ref{fig: g_q_plot}, while the curves for different sizes do intersect, they do not converge at a single crossing points. Instead, the crossing points shift to higher values as the lattice size increases, consistent with the established understanding that the two-dimensional EA model lacks a finite-temperature critical point.
Fig. \ref{fig: chi_q_plot} displays the susceptibility of the spin overlaps, $\chi(q)$, indicating that at low temperature, $\chi(q)$ exhibits a linear relationship with $\beta$ on a logarithmic scale.

\begin{figure}[ht]
\centering
\subfigure[]{
    \includegraphics[width=0.41\columnwidth]{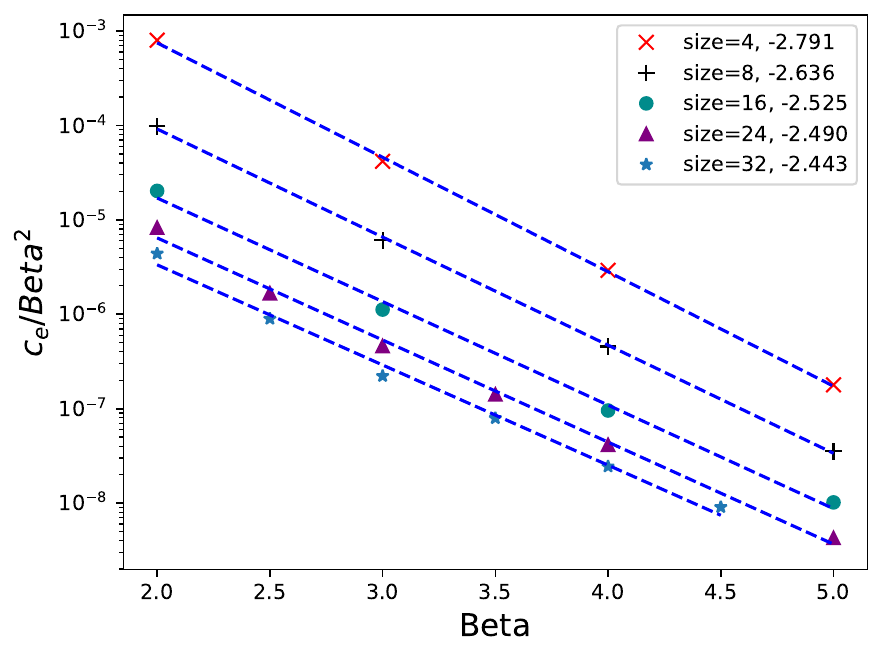}
    \label{fig: celog_lowt}
}
\subfigure[]{
    \includegraphics[width=0.48\columnwidth]{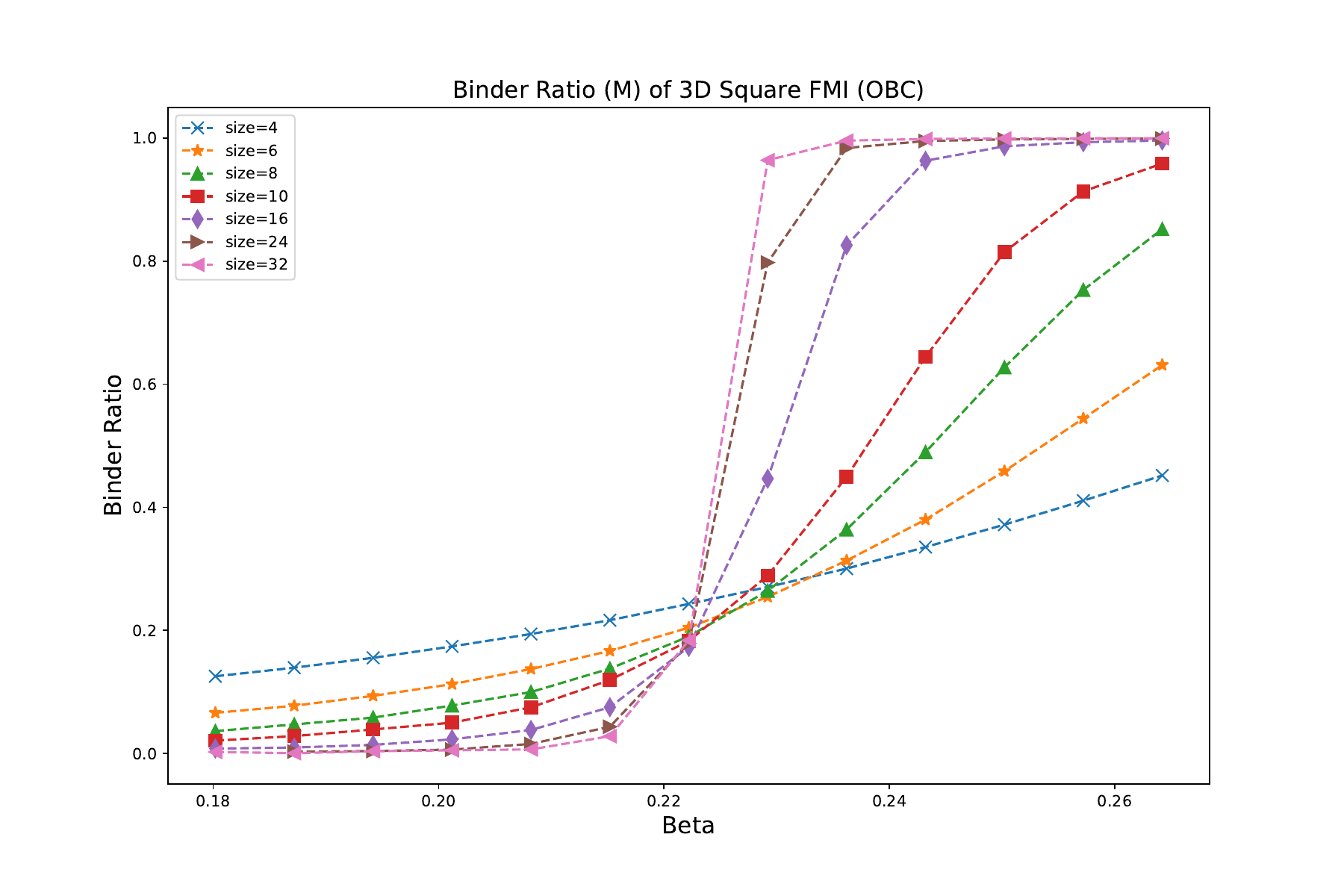}
    \label{fig: br_3dfmi}
}
\\
\subfigure[]{
    \includegraphics[width=0.48\columnwidth]{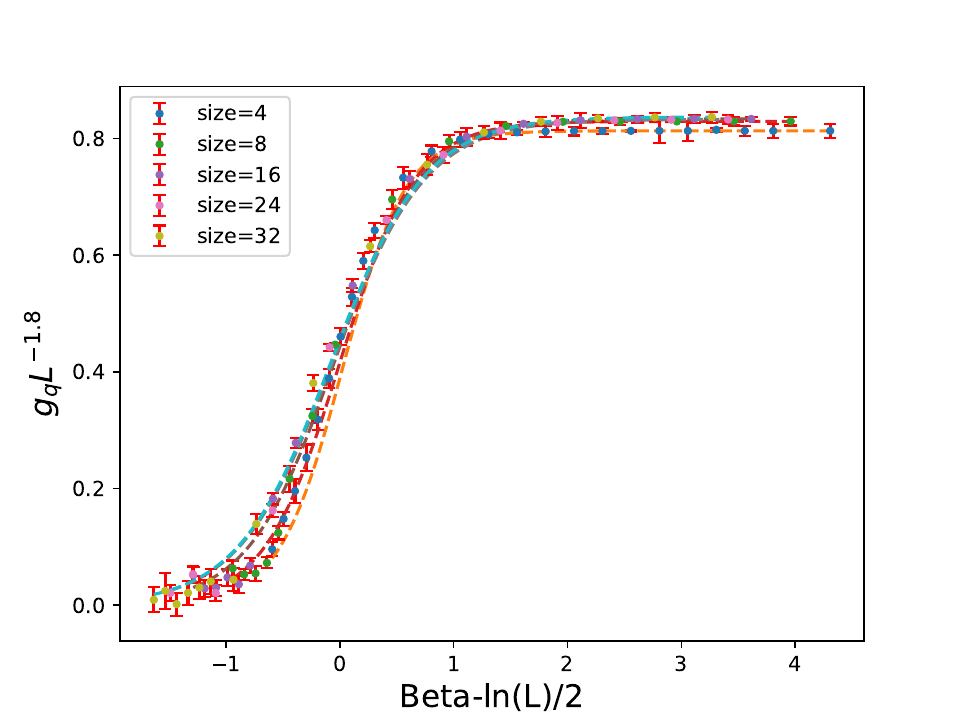}
    \label{fig: gq_scale}
}
\subfigure[]{
    \includegraphics[width=0.48\columnwidth]{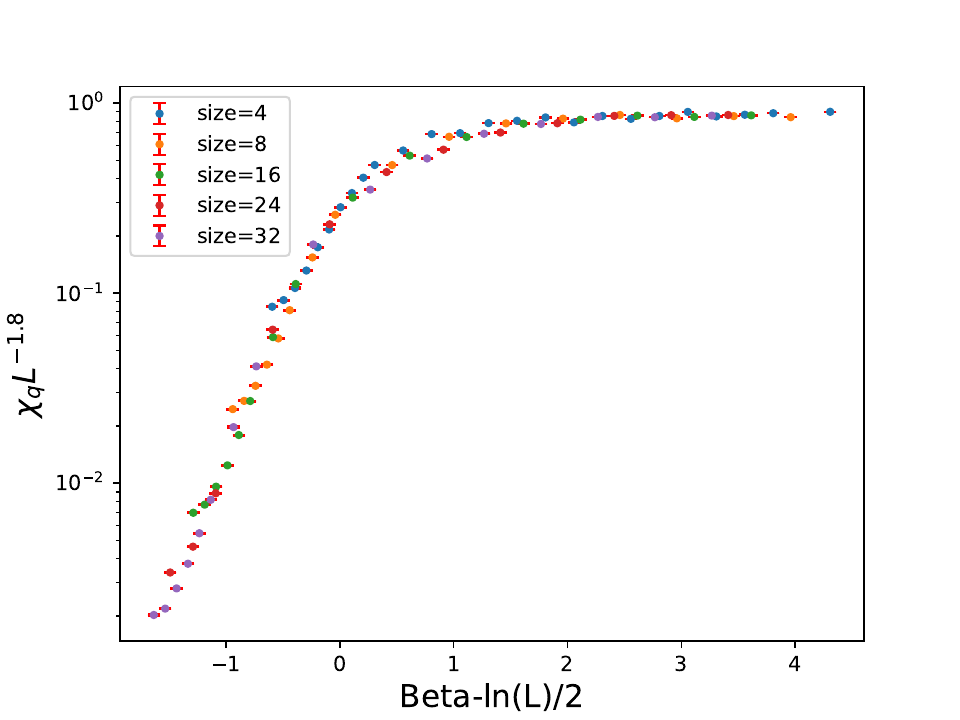} 
    \label{fig: chiq_scale}
}
\caption{(a) Specific heat $c_{e}$ of the 2D E-A model as a function of $\beta$ in the low low-temperature regime on a logarithmic scale, showing the expected scaling behavior at low temperatures.
(b) Binder ratio $g(m)$ of magnetisation as a function of beta for 3D Ising model.
(c) Binder ratio $g(q)$ as a function of the rescaled parameter $\beta - \ln(L) / 2$, demonstrating the collapse of data across different lattice sizes according to the proposed scaling rule.
(d) Rescaled susceptibility $\tilde{\chi}{(q)} = \chi(q)L^{-1.8}$ plotted against $\beta - \ln(L) / 2$, highlighting the consistency of data collapse with scaling predictions.}
\label{fig: scaling_law}
\end{figure}

It has been shown that the specific heat of the 2d EA model scales as $\beta^{2}e^{(-2\beta)}$ 
in the low-temperature regime in the thermodynamic limit \citep{wang2005worm}.
In Fig. \ref{fig: celog_lowt}, the scaling behavior of the specific heat $c_{e}$ with respect to lattice size at low temperatures approaches the aforementioned formula.
Fig. \ref{fig: gq_scale} and Fig. \ref{fig: chiq_scale} illustrate the data fitting for the Binder ratio $g(q)$ and susceptibility $\chi(q)$ at various lattice sizes using the scaling relations $g \sim \tilde{g}(\beta - \frac{1}{2}\ln{L})$, and $\chi \sim L^{2-\eta}\tilde{\chi}(\beta - \frac{1}{2}\ln{L})$, as proposed in \citep{saul1993exact, houdayer2001cluster}.
To demonstrate the flexibility of TGMH in simulating high-dimensional spin system, we apply it to the simulation of three-dimensional Ising model.
We obtain the Binder ratio of the magnetisation of the system as a function of temperature across different systems sizes,
ranging from $4$ to $32$, as shown in Fig. \ref{fig: br_3dfmi}.

\begin{figure}[ht]
\centering
\subfigure[]{
    \includegraphics[width=0.48\columnwidth]{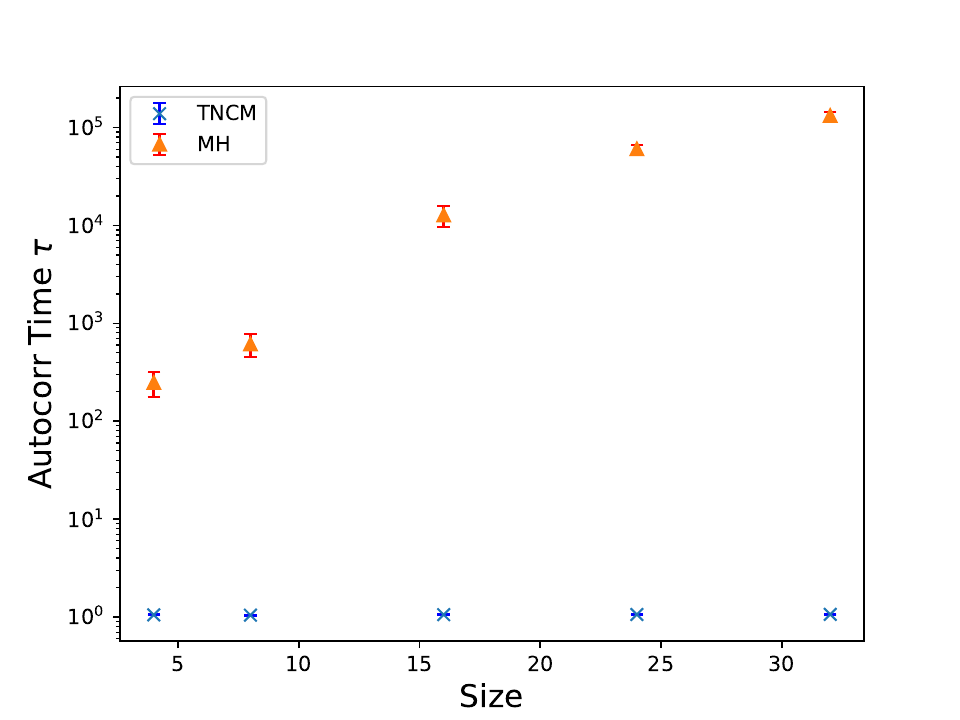}
    \label{fig: tau_size}
}
\subfigure[]{
    \includegraphics[width=0.48\columnwidth]{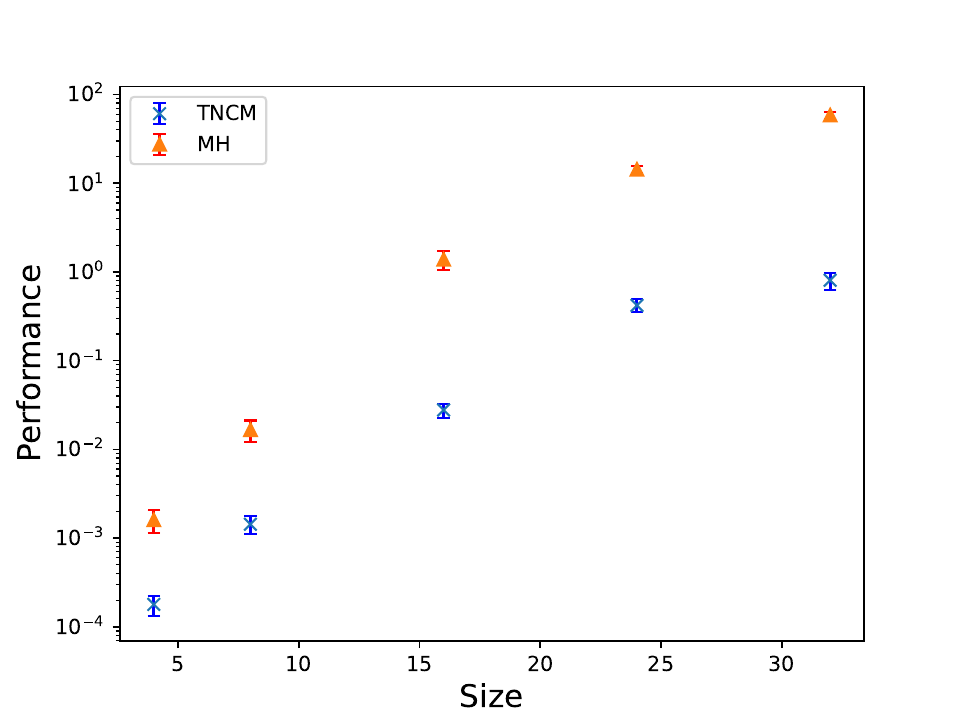} 
    \label{fig: perf_size}
}
\caption{(a) Autocorrelation Time $\tau$ of both TNCM and M-H method as a function of lattice size $L=4, 8, 16, 24, 32$ for $\beta=2.0$. (b) Performance metric $t_{0}\tau$ for both methods, using the same setup as in (a).}
\label{fig: perf_measure}
\end{figure}

We compare the performance of TNCM with the single spin-flipping Methropolis-Hasting (M-H) method using the quantity $\tau t_{0}$, which measures the real CPU time required for one effective sample\citep{swendsen1987nonuniversal}.
Here, $t_{0}$ and $\tau$ denote the real CPU time for a single MCMC sample, and the (integrated) autocorrelation time, respectively.
Fig. \ref{fig: tau_size} presents $\tau$ for both TNCM and M-H methods as a function of lattice size $L$.
The result shows that, as lattice size increases, the autocorrelation time for TNMC remains approximately constant around $1$, since the entire lattice is updated and accepted (with high probability) in each simulation step. 
In contrast, the autocorrelation time for M-H grows exponentially with lattice size.
In practical applications, both autocorrelation and real computational time must be considered.
Therefore, the performance measure $t_{0}\tau$ for different lattice sizes for both TNCM and M-H methods are illustrated in Fig. \ref{fig: perf_size}.
It can be observed that the CPU time required to obtain one effective sample using the TG formalism is reduced to about $1\%$ of that required by the single-spin flip algorithm.
In Appendix \ref{app: atc}, we 
present a theorem for constructing confidence intervals (error bars) for the statistics of EA models \citep{draper1995assessment}.


\vspace{-1px}
Although the simulations of small lattices are primarily presented in this work,
the TGMH method is well-suited for application to large-scale systems.
As the size of the system increases, making direct contraction of the corresponding TNs computationally expensive,
the advantages of the TGMH formalism become increasingly evident.
Unlike TNMH, TGMH enables the simulation of the entire system by dividing it into manageable clusters,
where TN contraction remains feasible within each cluster.
Optimizing cluster selection offers a potential avenue for future work.

\textbf{This work is dedicated to the memory of Prof. David Draper, a distinguished statistician whose lifelong contributions to scientific discovery and dedication to educating and mentoring future generations have left an enduring impact} \cite{in_memoriam_david}.

\newcommand{\noopsort}[1]{} \newcommand{\printfirst}[2]{#1}
  \newcommand{\singleletter}[1]{#1} \newcommand{\switchargs}[2]{#2#1}
\bibliography{reference.bib}

\clearpage

\appendix
\raggedbottom
\section{Computation of Conditional Distribution}
\label{app: contraction_cond}
For a cluster $C[\sigma]$ with spin configuration $\sigma_{1:n}$, edges $e_{1:k}$, boundary edges $e^{\prime}_{1:m}$ connecting the boundary spins with the spins within the cluster, and bond states $h_{1:k}$,
the conditional distribution of the cluster configuration $C[\sigma]$ given the boundary spin configuration $\partial{C}$ and the bond states can be expressed as follows:
\begin{align}
\label{eq: delta_chain}
    &p(C[\sigma_{1:n}]|\partial C, h_{1:k}) \propto \prod_{i=1}^{k}\delta(e_{i}, h_{i})\prod_{j=1}^{m}v(e^{\prime}_{j}), \\
    &\delta{(e_{i}, h_{i})} = \delta{(\sigma^{(e_{i})}_{1}, \sigma^{(e_{i})}_{2})}I[h_{i} = 1] + \delta{(\sigma^{(e_{i})}_{1}, -\sigma^{(e_{i})}_{2})}I[h_{i} = 2], \\
    &v(e_{j}^{\prime}) = \delta{(\sigma^{(e^{\prime}_{j})}_{1}, \sigma^{(e^{\prime}_{j})}_{2})}X + \delta{(\sigma^{(e^{\prime}_{j})}_{1}, -\sigma^{(e^{\prime}_{j})}_{2})}Y,
\end{align}
$(\sigma_{1}^{(e_{i}}, \sigma_{2}^{(e_{i}})$ represents the spin pairs located at the two ends of the edge $e_{i}$.
It can be demonstrated that only two spin configurations within the cluster, 
$C[\sigma]$ and $C[\bar{\sigma}]$ (where $C[\bar{\sigma}]$ represents the configuration with all spins in the cluster flipped), have non-zero probability.
This can be established by contradiction:
Suppose there exists another cluster configuration $C[\sigma^{\prime}]$ with non-zero probability,
distinct from both $C[\sigma]$ and $C[\bar{\sigma}]$.
In such a case, we can divide the spins into two groups: one group consists of spins that remain unchanged compared to $C[\sigma]$,
and the other group consists of spins that are flipped relative to $C[\sigma]$.
Since all spins belong to the same cluster, there must be at least one edge connecting a spin from the unchanged group to a spin in the flipped group.
However, this would imply a zero probability for the configuration $C[\sigma^{\prime}]$,
 which contradicts the assumption that $C[\sigma^{\prime}]$ has non-zero probability.
 Thus, we have proven that only two configurations of the cluster,
 $C[\sigma]$ and $C[\bar{\sigma}]$ can have non-zero probability.
Moreover, based on Eq. \ref{eq: delta_chain},
the probability ratio of the conditional distributions for $C[\sigma_{1:n}$ and $C[\bar{\sigma}_{1:n}]$ can be expressed as 
\begin{align}
    \frac{p(C_{i}[\sigma]|C_{i}[\mathbf{h}], \partial{C_{i}}[\sigma])}{p(C_{i}[\bar{\sigma}]|C_{i}[\mathbf{h}], \partial{C_{i}}[\sigma])} = (\frac{X}{Y})^{n-m}.
\end{align}

\section{TG with Ghost Field Method}
\label{app: ghost_tensor}
Similarly to the construction of the auxiliary variable for the Boltzmann matrix $B$,
we define the following relationship for $C_{ij}$,
\begin{equation}
\label{eq: c_ij}
    C_{ij} = \sum_{l=1}^{3}C^{l}_{ij},
\end{equation}
where
$C_{ij}^{1} = 
\begin{pmatrix}
    e^{\tilde{B}} - E & 0 \\
    0 & e^{\tilde{B}} - E
\end{pmatrix}$,
$C_{ij}^{2} = 
    \begin{pmatrix}
    0 & e^{-\tilde{B}} - F \\
    e^{-\tilde{B}} - F & 0 
    \end{pmatrix}$,
and
$C_{ij}^{3} = 
\begin{pmatrix}
   E & F \\
   F & E
\end{pmatrix}$,
and the conditional distribution of the auxiliary variables $\mathbf{t}^{(ij)}$ for matrix $C_{ij}$ is as,
\begin{align}
    &p(\mathbf{t}^{(ij)} = (1, 0, 0)|\sigma_{i}, \sigma_{j}) = (1 - Ee^{-\tilde{B}})\delta_{\sigma_{i}, \sigma_{j}}, \\
    &p(\mathbf{t}^{(ij)} = (0, 1, 0)|\sigma_{i}, \sigma_{j}) = (1 - Fe^{\tilde{B}})\delta_{\sigma_{i}, \sigma_{j}}, \\
    &p(\mathbf{t}^{(ij)} = (0, 0, 1)|\sigma_{i}, \sigma_{j}) = 
    Ee^{-\tilde{B}}\delta_{\sigma_{i}, \sigma_{j}} + Fe^{\tilde{B}}\delta_{\sigma_{i}, -\sigma_{j}}.
\end{align}
The edge $(i, j)$ in above formula refers to a "ghost" edge, indicating that one end of the edge $(i, j)$ always connects to the ghost spin,
with $\sigma_{i}=\sigma_{\text{ghost}}$.
The conditional distribution of the spin configuration of $C[\sigma]$ given the bond states and also the boundary spins can then be expressed as
\begin{align}
    &p(C[\sigma_{1:n}]|\partial C, h_{1:k}) \propto \prod_{i=1}^{k}\delta(e_{i}, h_{i})\prod_{j=1}^{m}v(e^{\prime}_{j})
    \prod_{l=1}^{o}\delta(e_{l}, t_{l})
    \prod_{n=1}^{p}u(e^{\prime}_{n}), \\
    &\delta{(e_{i}, h_{i})} = \delta{(\sigma^{(e_{i})}_{1}, \sigma^{(e_{i})}_{2})}I[h_{i} = 1] + \delta{(\sigma^{(e_{i})}_{1}, -\sigma^{(e_{i})}_{2})}I[h_{i} = 2], \\
    &\delta{(e_{i}, t_{i})} = \delta{(\sigma^{(e_{i})}_{1}, \sigma^{(e_{i})}_{2})}I[t_{i} = 1] + \delta{(\sigma^{(e_{i})}_{1}, -\sigma^{(e_{i})}_{2})}I[t_{i} = 2], \\
    &v(e_{j}^{\prime}) = \delta{(\sigma^{(e^{\prime}_{j})}_{1}, \sigma^{(e^{\prime}_{j})}_{2})}X + \delta{(\sigma^{(e^{\prime}_{j})}_{1}, -\sigma^{(e^{\prime}_{j})}_{2})}Y, \\
    &u(e_{j}^{\prime}) = \delta{(\sigma^{(e^{\prime}_{j})}_{1}, \sigma^{(e^{\prime}_{j})}_{2})}E + \delta{(\sigma^{(e^{\prime}_{j})}_{1}, -\sigma^{(e^{\prime}_{j})}_{2})}F.
\end{align}
Applying the same reasoning as in the previous proof,
the conditional probability ratio is obtained as
\begin{align}
    \frac{p(C_{i}[\sigma, \sigma_{\text{ghost}}]|C_{i}[\mathbf{h}, \mathbf{t}], \partial{C_{i}}[\sigma, \sigma_{\text{ghost}}])}{p(C_{i}[\bar{\sigma}, \bar{\sigma}_{\text{ghost}}]|C_{i}[\mathbf{h}, \mathbf{t}], \partial{C_{i}}[\sigma, \sigma_{\text{ghost}}])} = (\frac{X}{Y})^{n-m}(\frac{E}{F})^{p-q},
\end{align}
Here, $p$ and $q$ represent the numbers of ghost edges (i.e., edges connecting the cluster spin and boundary spin) that have parallel and anti-parallel spin pairs, respectively.
It is evident that setting $X=Y=e^{-K}$ and $E=F=e^{-\tilde{B}}$ yields the ghost S-W algorithm.

In following context, we explain the measurement of a quantity $M$ in the ghost TG formalism.
\begin{align}    
    \mathbb{E}[M(\sigma_{1:n})] &= 
\frac{\sum_{\sigma_{1:n}}e^{-\beta E(\sigma_{1:n})}M(\sigma_{1:n})}{\sum_{\sigma_{1:n}}e^{-\beta E(\sigma_{1:n})}} \\
&= \frac{\sum_{\sigma_{1:n}}e^{-\beta \tilde{E}(\sigma_{1:n}, \sigma_{\text{ghost}}=1)}\tilde{M}(\sigma_{1:n}, \sigma_{\text{ghost}}=1)}{\sum_{\sigma_{1:n}}e^{-\beta \tilde{E}(\sigma_{1:n}, \sigma_{\text{ghost}}=1)}}\\
&= \frac{\sum_{\sigma_{\text{ghost}}
\in \mathbf{G}
}\sum_{\sigma_{1:n}}e^{-\beta \tilde{E}(\sigma_{1:n}, \sigma_{\text{ghost}})}\mu(\sigma_{\text{ghost}})\tilde{M}(\sigma_{1:n}, \sigma_{\text{ghost}})}{\sum_{\sigma_{\text{ghost}}\in \mathbf{G}}\sum_{\sigma_{1:n}}e^{-\beta \tilde{E}(\sigma_{1:n}, \sigma_{\text{ghost}})}\mu(\sigma_{\text{ghost}})} \\
&= \mathbb{E}[\tilde{M}(\sigma_{1:n}, \sigma_{\text{ghost}})],
\end{align}
where $\tilde{M}[\sigma_{1:n}, \sigma_{\text{ghost}}]$ is defined as $M[\sigma_{\text{ghost}}*\sigma_{1:n}]$,
and $\mu(\sigma_{\text{ghost}})$ represents the measure of the elements.
In this proof, we consider the discrete symmetry group $\mathbf{G}$ of the simulating system,
with the ghost spin evaluated by the elements of this symmetry group.

\section{Asymptotics of Measurements in E-A Models via Monte Carlo Simulation}
\label{app: atc}
In the context of the E-A model,
the Monte Carlo simulation involves generating $N$ realizations of disorder,
denoted as $\mathbf{J}_{1:N}$,
where each $\mathbf{J}_{i}$ is independently drawn from a distribution $p(\mathbf{J}_{i})$,
typically either binomial or Gaussian.
For each disorder realization,
$M$ samples $\underline{\sigma}^{1:N}$ are obtained,
with $\underline{\sigma}^{i}$ following the Boltzmann distribution $p_{\text{Boltzmann}}(\cdot)$.

Given this setup, we present the following theorem concerning the asymptotic behavior of the Monte Carlo estimator for the measurement $\mathbb{E}_{\text{J}, T}[A]$ of quantity $A$.

\begin{theorem}[Asymptotics for the MC estimator of E-A models]
Consider the simulation data for the quantity $A$ in the E-A model,
represented as $\{A_{ij}, i=1, \cdots, M, j=1, \cdots, N\}$,
where
\begin{align*}
    &A_{ij} \overset{\text{i.i.d.}}{\sim} p_{\text{Boltzmann}}(A_{ij}|\mathbf{J}_{i}), \quad j = 1, \cdots, M,\\
    &\mathbf{J}_{i} \overset{\text{i.i.d.}}{\sim} p(\mathbf{J}_{i}), \quad i = 1, \cdots, N.
\end{align*}
we have following asymptotic result as,
\begin{align*}
\lim_{N\to\infty}\lim_{M\to\infty}\frac{1}{\sqrt{N}}\sum^{N}_{i=1}[\frac{1}{\sqrt{M}\sigma_{i}}\sum^{M}_{j=1}A_{ij}] \overset{\text{Dist}}{\to} N(\mu, 1),
\end{align*}
where $\sigma_{i} = \mathbb{E}_{\text{T}}[(A_{i} - \mathbb{E}_{\text{T}}[A_{i}])^{2}]$ represents the second raw moment of the random variable $A_{i}$ for the $i$-th realization of disorder $\mathbf{J}_{i}$.
\end{theorem}
This theorem facilitates the construction of a confidence interval (CI) for the MC estimator $\hat{\mu} = \frac{1}{NM}\sum_{ij}A_{ij}$ of $\mu$ with a significance level $\alpha$.
The CI can be expressed as 
$(\hat{\mu} - \hat{\sigma}\Phi^{-1}(\frac{\alpha}{2}), \,\hat{\mu} + \hat{\sigma}\Phi^{-1}(1-\frac{\alpha}{2})$ \cite{davids_notes},
where
$\hat{\sigma} = \frac{1}{N}\sqrt{\sum_{i}\hat{\sigma}_{i}^{2}\frac{1+\rho_{i}^{1}}{1-\rho_{i}^{1}}}$ is an estimate of the standard error,
with $\rho^{(1)}_{i}$ representing the first-order autocorrelation in the $i$-th simulation.
Here, $\hat{\sigma}^{2}_{i}$ is the sample variance of the simulation data $\{A_{ij}, \, j=1, \cdots, N\}$ for the $i$-th realization $\mathbf{J}_{i}$ of the disorder.
The function $\Phi^{-1}(\cdot)$ represents the inverse of the Cumulative Distribution Function (CDF) of the standard normal distribution.

\section{Detailed Sampling Procedure by TGMH}
\label{app: tncm_tensors}
\begin{figure}[ht]
\centering
\subfigure[]{
    \includegraphics[width=0.48\columnwidth]{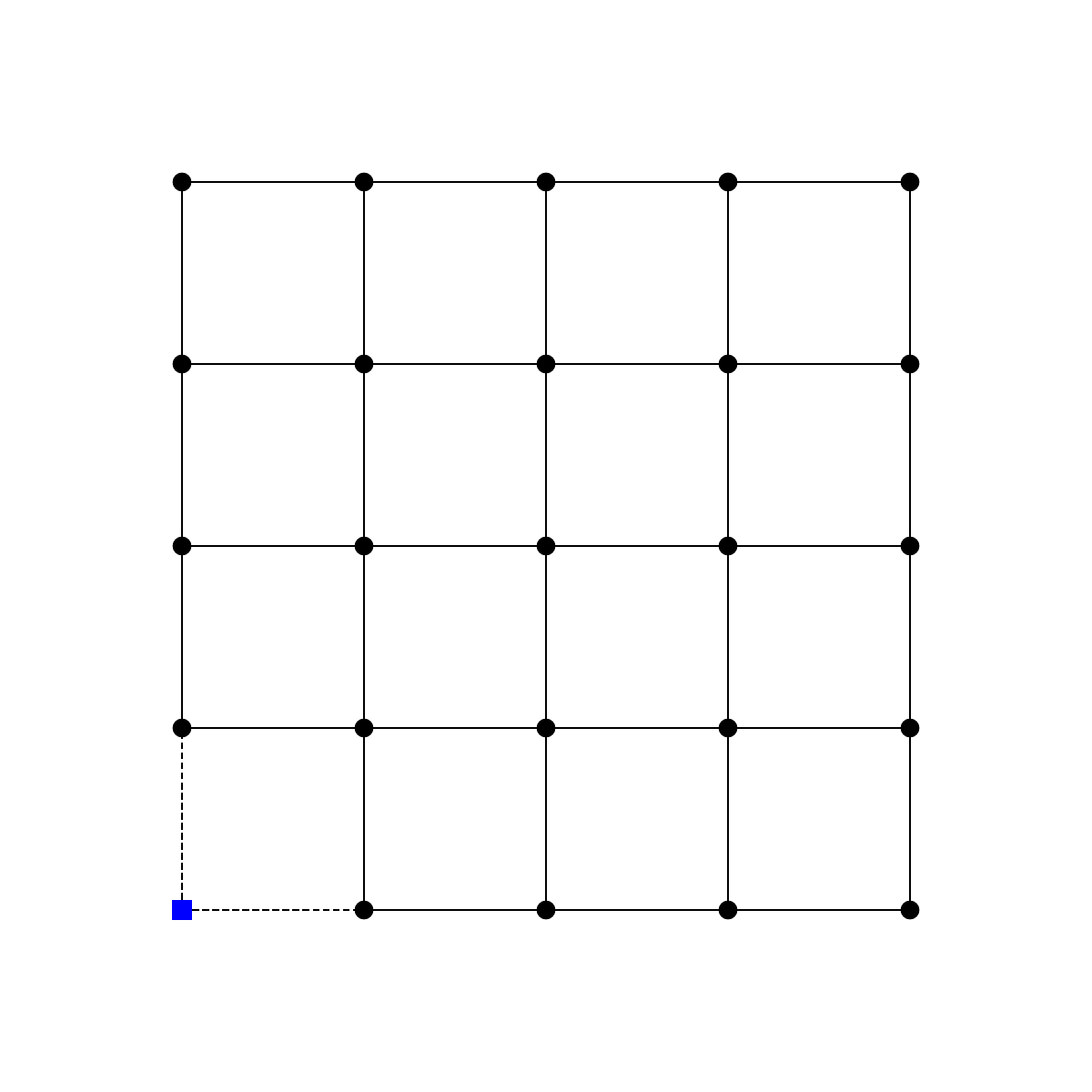}
    \label{fig: tnmc_1}
}
\subfigure[]{
    \includegraphics[width=0.48\columnwidth]{./figs/tnmc_step_2.pdf}
    \label{fig: tnmc_2}
}\\
\subfigure[]{
    \includegraphics[width=0.48\columnwidth]{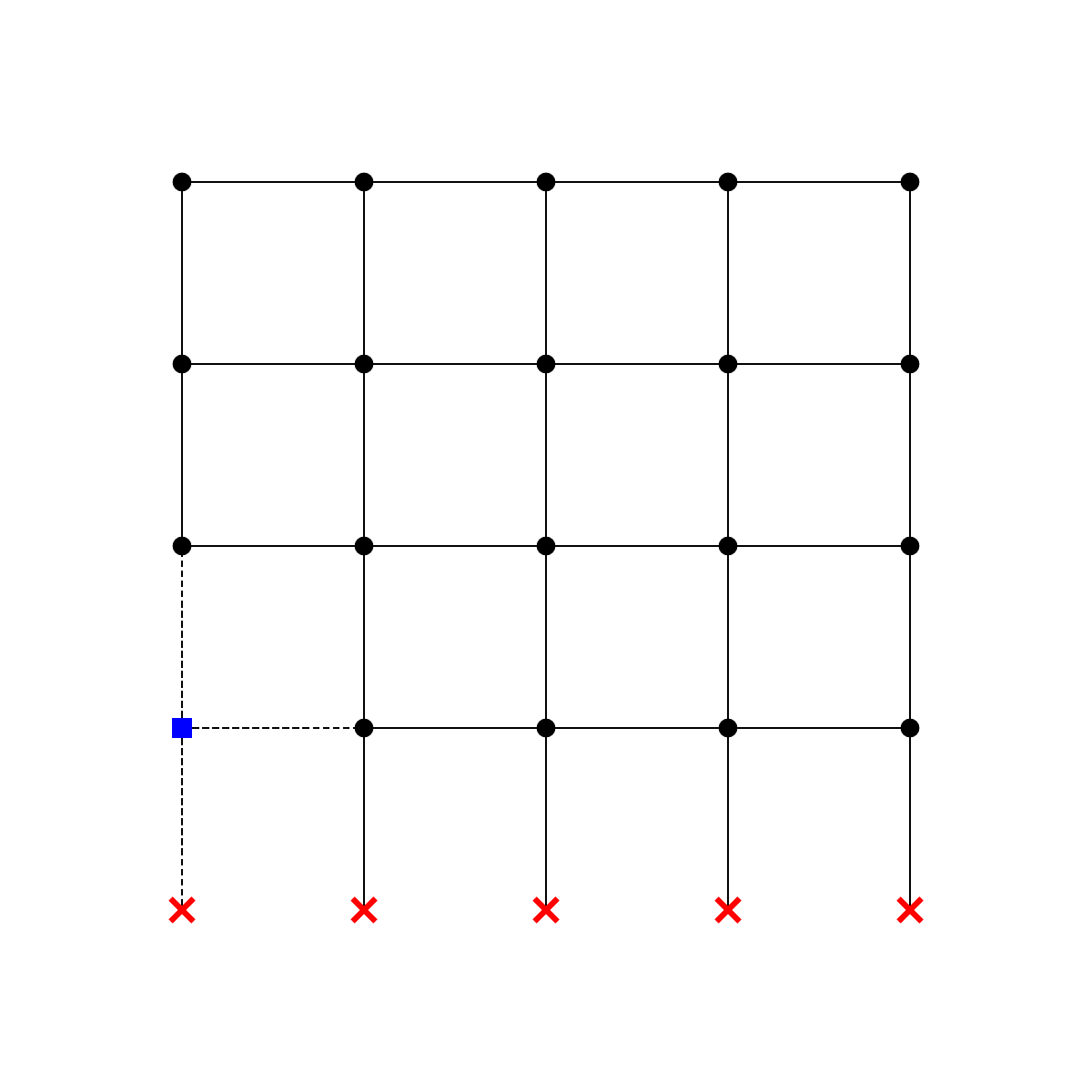}
    \label{fig: tnmc_3}
}
\subfigure[]{
    \includegraphics[width=0.48\columnwidth]{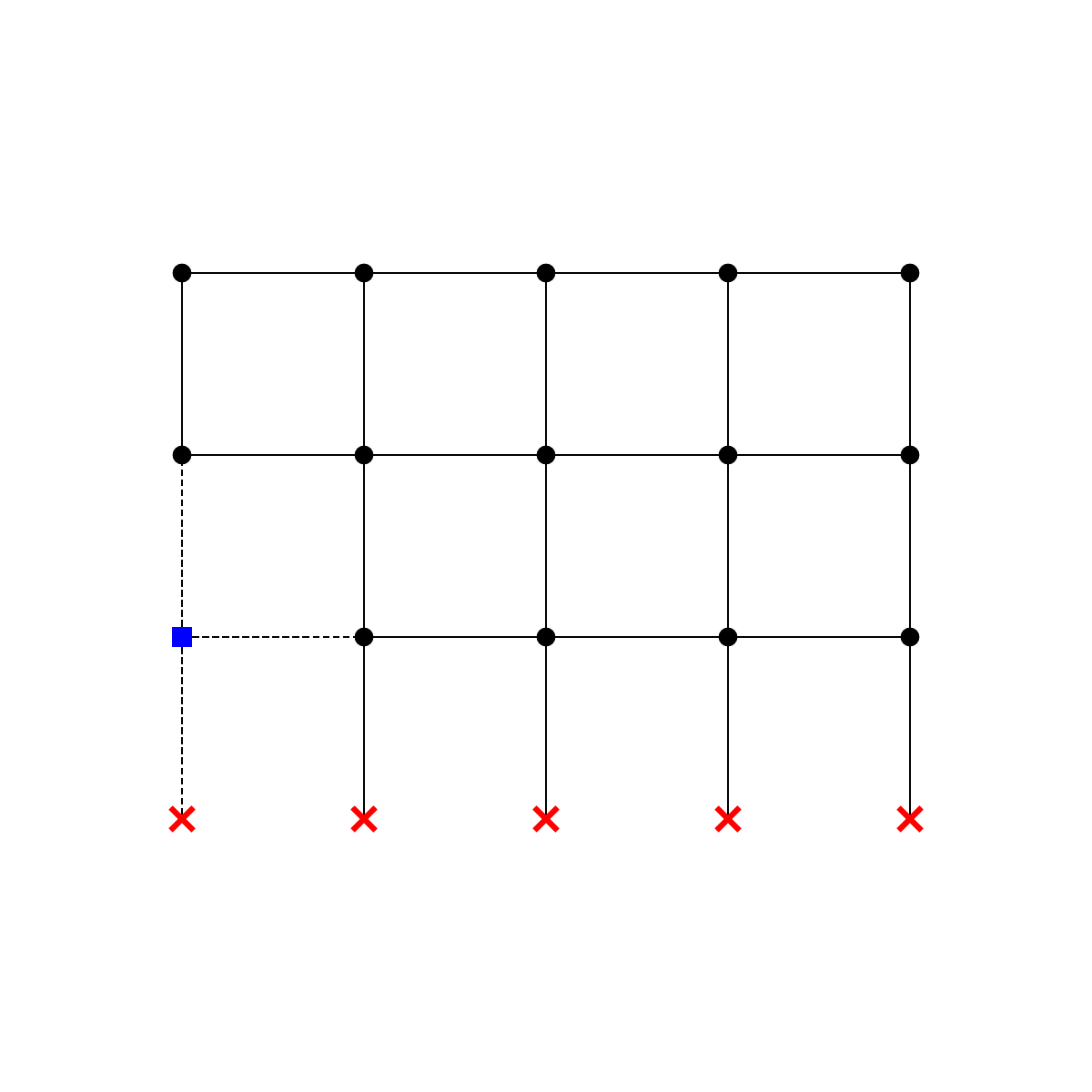}
    \label{fig: tnmc_4}
}
\caption{From (a) to (d), sampling procedure is illustrated as iterating through all spins from left to right and from bottom to the top of lattice.
(a) demonstrates the computation of the proposal distribution $\tilde{p}(\sigma_{1})$, which is used to sample the proposed spin for first site $\sigma_{1}^{\prime}$.
(b) shows the computation of the proposal distribution $\tilde{p}(\sigma_{4}|\sigma_{1:3})$, where the previously sampled $\sigma_{1:3}$ are treated as external fields.
(c) illustrates the sampling of the proposed spin at the first column, second-to-last row, with the spins of the last row treated as external fields.
(d) depicts the sampling of the proposed spin at the first column, third-to-last row are considered as external fields due to the Markovian properties of the square lattice structure.}
\label{fig: tnmc}
\end{figure}
In this section, we explain the procedure of sampling from the proposal distribution
$\tilde{p}(\text{C}(\sigma)|\partial{\text{C}}[\sigma])$
using tensor contraction.
As all spins on the boundary $C[\sigma]$ can be treated as external field,
we will illustrate the sampling process of the entire lattice 
$\hat{p}(\sigma_{1:n})$
as illustrated in Fig. \ref{fig: tnmc}.
The central idea, as discussed in previous context, involves sampling from the conditional distributions,
and subsequently computing the proposal distribution for a specific configuration $\sigma_{1:n}$ using chain rule: 
$\tilde{p}(\sigma_{1})\prod_{i=2}^{n}\tilde{p}(\sigma_{i}|\sigma_{1:i-1})$.
For a given state of the spin lattice,
the conditional distribution represented by tensor contraction is expressed as
\begin{align}
\label{eq: condition_vector}
p(\sigma_{i}|\sigma_{1:i-1}) = p^{u} \propto \sum_{\{e\}}T^{(2)}_{ij} \cdots T^{(3)}_{klm} \cdots T^{(4)}_{nopq}\cdots v^{(1)}_{r} \cdots S^{u}_{x \cdots z},
\end{align}
In this equation, $T_{ij}^{(2)}$ denotes the tensors with two edges, such as those located at the top left, top right and bottom right sites in Fig. \ref{fig: tnmc_1}.
$T^{(3)}_{ijk}$ refers to the three-way tensors located along the top row or bottom row (except the corner),
while $T^{(4)}_{ijkl}$ represents the tensor with four edges.
The vector $v^{(1)}_{i}$ is an external field vector associated with the red cross site in Fig. \ref{fig: tnmc}.
Notably, a single external field (red cross site) may have multiple external field vectors,
as it can interact with multiple spins.
The tensor $S^{u}_{i\cdots j}$ corresponds to the spin site currently being sampled, represented by the blue square with dashed edges in Fig \ref{fig: tnmc}.
The number of indices for this tensor is determined by the number of edges at the site plus one additional sampling index $u$,
which retains the probability of each spin state rather than summing them.
The sampling index $u$ is denoted as a superscript index, as it is not contracted,
while all other subscript indices, represented by the set of duplicated edges $\{e\}$ in Eq. \ref{eq: condition_vector}, are contracted.
The position (up or down) of the indices in $S^{u}_{i, \cdots, j}$ does not indicate any distinction.
The definitions of the tensors in Eq. \ref{eq: condition_vector} are as follows,
\begin{align*}
    &T^{(2)}_{ij} = \sum_{kl}\sqrt{B}_{ik}\sqrt{B}_{jl}I_{kl},\\
    &T^{(3)}_{ijk} = 
    \sum_{mln}\sqrt{B}_{im}\sqrt{B}_{jl}\sqrt{B}_{kn}I_{mln},\\
    &T^{(4)}_{ijkl} = \sum_{mnpq}\sqrt{B}_{im}\sqrt{B}_{jn}\sqrt{B}_{kp}\sqrt{B}_{lq}I_{mnpq}, \\
    &v^{(1)}_{i}(\sigma) = \sum_{j}\sqrt{B}_{ij}L_{j}, \quad L_{j} = [\delta{(\sigma, \uparrow)}, \delta{(\sigma, \downarrow)}], \\
    &S^{u}_{i\cdots j} = \sum_{k, \cdots, l}\sqrt{B}_{ik}\cdots \sqrt{B}_{jl}I_{k, \cdots, l}^{u}
\end{align*}
where $I_{kl} = \delta_{kl}$, $I_{mln} = \delta_{ml}\delta_{ln}$, $I_{mnpq} = \delta_{mn}\delta_{np}\delta_{pq}$,
and $I^{u}_{k, \cdots, l} = \delta_{u, k}\delta_{k, \cdot}\cdots\delta_{\cdot, l}$.
The matrix $\sqrt{B}_{ij}$ satisfies the condition that $\sum_{l}\sqrt{B}_{il}\sqrt{B}_{lj} = B_{ij}$.
By exactly contracting all duplicated indices in Eq. \ref{eq: condition_vector},
while leaving the "sampling" index $u$ uncontracted,
and subsequently normalizing the $L^{1}$ norm of the resulting vector,
the conditional probability vector $p(\sigma_{k}|\sigma_{1:k-1})$ is obtained.
However, exact contraction is computationally infeasible for large tensor networks,
In such cases, bond dimension truncation is applied to approximate the contraction,
yielding the proposal probability vector $\tilde{p}(\sigma_{k}|\sigma_{1:k-1})$. 
In practice, boundary MPS contraction methods can be employed to perform these computations efficiently,
along with caching certain intermediate contraction results to further reduce computational costs.
\end{document}